\documentclass{ws-ijmpa}
\usepackage{slashed,feynmf,epsfig,verbatim,cite}
\renewcommand{\vec}[1]{\boldsymbol{\mathrm{#1}}}

\begin{document}

\title{A finite electroweak model without a Higgs particle}

\author{J. W. Moffat$^{\dag *}$ and V. T. Toth$^{\dag}$\\~\\
$^\dag$Perimeter Institute, 31 Caroline St North, Waterloo, Ontario N2L 2Y5, Canada\\
$^*$Department of Physics, University of Waterloo, Waterloo, Ontario N2L 3G1, Canada}

\maketitle

\begin{abstract}
An electroweak model in which the masses of the $W$ and $Z$ bosons and the fermions are generated by quantum loop graphs through a symmetry breaking is investigated. The model is based on a regularized quantum field theory in which the quantum loop graphs are finite to all orders of perturbation theory and the massless theory is gauge invariant, Poincar\'e invariant, and unitary. The breaking of the electroweak symmetry $SU_L(2)\times U_Y(1)$ is achieved without a Higgs particle. A fundamental energy scale $\Lambda_W$ (not to be confused with a naive cutoff) enters the theory through the regularization of the Feynman loop diagrams. The finite regularized theory with $\Lambda_W$ allows for a fitting of low energy electroweak data. $\Lambda_W\sim 542$~GeV is determined at the $Z$ pole by fitting it to the $Z$ mass $m_Z$, and anchoring the value of $\sin^2\theta_w$ to its experimental value at the $Z$ pole yields a prediction for the $W$ mass $m_W$ that is accurate to about $0.5\%$ without radiative corrections. The scattering amplitudes for $W_LW_L\rightarrow W_LW_L$ and $e^+e^-\rightarrow W_L^+W_L^-$ processes do not violate unitarity at high energies due to the suppression of the amplitudes by the running of the coupling constants at vertices. There is no Higgs hierarchy fine-tuning problem in the model. The unitary tree level amplitudes for $W_LW_L\rightarrow W_LW_L$ scattering and $e^+e^-\rightarrow W_L^+W_L^-$ annihilation, predicted by the finite electroweak model are compared with the amplitudes obtained from the standard model with Higgs exchange. These predicted amplitudes can be used to distinguish at the LHC between the standard electroweak model and the Higgsless model.
\end{abstract}

%\pacs{11.10.Lm,11.15.Ex,12.15.-y,12.60.Cn}

%11.10.Lm    Nonlinear or nonlocal theories and models
%11.15.Ex    Spontaneous breaking of gauge symmetries
%12.15.-y    Electroweak interactions
%12.60.Cn    Extensions of electroweak gauge sector
%13.66.Fg    Gauge and Higgs boson production in e-e+ interactions
%14.70.Fm    W bosons

%\maketitle

\begin{fmffile}{nhfigs}

\fmfcmd{%
 vardef bar (expr p, len, ang) =
  ((-len/2,0)--(len/2,0))
     rotated (ang + angle
       direction length(p)/2 of p)
       shifted point length(p)/2 of p
 enddef;
 style_def crossed_phantom expr p =
  draw_phantom p;
  ccutdraw bar (p, 3mm,  60);
  ccutdraw bar (p, 3mm, -60)
 enddef;}

\section{Introduction}

In previous work, a finite electroweak (FEW) model was developed based on a quantum field theory which is finite to all orders of perturbation theory~\cite{Moffat1991,Clayton1991b,Clayton1991a,Moffat2007f}. For massless particles, the model is gauge invariant under an extended gauge invariance, which contains $SU_L(2)\times U_Y(1)$. All tree graphs are strictly local and point like, so that the classical theory does not violate macrocausality. On the other hand, the quantum loop graphs are finite due to the nonlocal field operators in the interaction Lagrangian. The quantum field theory is based on a regularized UV complete field theory, which for massless particles is gauge invariant, Poincar\'e invariant, and unitary \cite{Moffat1991,Clayton1991b,Clayton1991a,Moffat2007f,Moffat1989,Moffat1990,Evens1991,Moffat1991b,Hand1991,Kleppe1991,Kleppe1992,Cornish1992b,Cornish1992c,Cornish1992,Kleppe1993,Clayton1994,Paris1995,Paris1996,Saini1997,Efimov1967,
Efimov1968,Efimov1966,Efimov1967b}.

The standard electroweak (EW) model gains mass for the $W$ and $Z$ bosons, while keeping the photon massless by introducing a scalar field into the classical action. This scalar degree of freedom is assumed to transform as an isospin doublet, spontaneously breaking the $SU_L(2)\times U_Y(1)$ by a Higgs mechanism~\cite{Higgs1964a,Higgs1964b,Englert1964,Guralnik1964,Higgs1966,Glashow1961,Weinberg1967,Salam1968} at the purely classical tree graph level. The predicted Higgs particle has not been detected in high energy experiments. Recent results at the Tevatron accelerator show that given the very precise value of the top quark mass, $m_t=171.2\pm 2.1$~GeV (correct to $1.2\%$ \cite{PDG2008}) the accurate mass of the $W$ meson, $m_W=80.398\pm 0.025$~GeV \cite{PDG2008}, and that the standard EW model is correct (without additional undetected particles), then the Higgs boson must be light with a mass less than $150$~GeV. Preliminary results from the Tevatron experiments have not detected the Higgs particle, but the LHC with its 14~TeV energy and larger luminosity will hopefully settle the issue as to whether the Higgs particle exists in nature.

The origin of the symmetry breaking mechanism remains elusive after almost 50 years. The standard and commonly accepted explanation is a spontaneous symmetry breaking framework in which the symmetry $SU_L(2)\times U_Y(1)$ is not broken by the interactions but is ``softly'' broken by the asymmetry of the ground state or the vacuum state. For a global (spacetime independent) symmetry the spontaneously broken gauge directions give rise to massless, spin-zero scalar Nambu-Goldstone bosons. For broken gauge directions corresponding to a spacetime dependent local symmetry, the Nambu-Goldstone bosons associate with the $W$ and $Z$ gauge bosons to form the massive $W$ and $Z$ gauge bosons. The initially massless gauge bosons have two transverse polarization states that are given, in a comoving frame, by the vector:
\begin{equation}
\label{transverse}
\epsilon^\mu_\pm=\frac{1}{\sqrt{2}}(0,1,\pm i,0),
\end{equation}
where $\mu=0,1,2,3$ and the $\hat{z}$ or 3-direction is pointed along the direction of motion. In a frame in which the massive gauge boson moves in the ${\hat{z}}$ direction, the two transverse spin states are given by Eq. \ref{transverse}, and the third spin state is determined by the longitudinal polarization vector:
\begin{equation}
\label{longitudinal}
\epsilon^\mu_0=\frac{1}{m_V}(|\vec{p}|,0,0,E),
\end{equation}
where $m_V$ is the mass of the gauge boson and $\vec{p}$ and $E$ denote the three-momentum and energy, respectively.

In a theory with a spontaneously broken symmetry, such as the standard EW model with a Higgs meson, an equivalence theorem can be proved~\cite{Peskin1995,Burgess2007}. At energies large compared to the gauge boson mass $m_V$, the longitudinal mode can be identified with the underlying Nambu-Goldstone scalar boson produced in the symmetry breaking sector. The three longitudinal gauge boson modes $W_L^\pm$ and $Z_L$ are identified with the three scalar Nambu-Goldstone modes $w^\pm=\sqrt{2}(w_1\pm iw_2)$ and $z=w_3$. It is therefore important to study the longitudinally polarized $W_L$ and $Z_L$ at the LHC, so that we can discover the dynamics of the symmetry breaking mechanism.

The strong theoretical prejudice in favor of the Higgs spontaneous symmetry breaking mechanism, despite the lack of firm experimental confirmation, is based on the renormalizability of the standard EW model~\cite{tHooft1971a,tHooft1971b,tHooft1973}. The renormalizability criterion is intimately connected to the cancelation of bad high energy behavior of the longitudinal components of the gauge bosons that arises because of the high-energy behavior of (\ref{longitudinal}). This criterion of renormalizability of the EW model runs into obstacles because of the experimentally detected masses of the neutrinos. Incorporating a massive Dirac neutrino into the standard model requires dimensionless coupling constants of order $10^{-11}$ or less. The alternative of a Majorana neutrino leads to a violation of lepton number conservation. Additional possibilities, such as using higher-dimensional interactions to account for neutrino oscillations, or extending the standard model particle content with new, not yet observed particles are equally problematic \cite{Burgess2007}. This is a direct failure of the minimal standard EW model. If there are no further undetected particles to be discovered at the LHC, then we must seriously consider a new kind of quantum field theory framework which is generically finite and does not rely on some {\it ad hoc} renormalizability criterion. This is one of the motivations for investigating a finite quantum field theory and basing a different EW model on such a finite QFT formalism~\cite{Moffat1991,Clayton1991b,Moffat2007f}.

Alternatives to the standard EW model that perform the task of the minimal Higgs sector in giving masses to the $W$ and $Z$ bosons include supersymmetric models, or dynamical symmetry breaking models such as the strong interaction class of technicolor models~\cite{Burgess2007}. As we shall find in the following, and as described in ~\cite{Moffat1991,Clayton1991b,Moffat2007f}, there are indeed models of the EW symmetry breaking based on a finite QFT that can claim theoretical and experimental success, and there may be other models that we have not yet imagined.

A significant problem with the standard EW model based on a Higgs mechanism is the instability of the Higgs particle mass $m_H$. The lowest order Higgs mass self-interaction is quadratically divergent and produces a severe mass hierarchy problem that has plagued the standard model from the beginning. Efforts such as the ``little Higgs'' model must postulate undetected particles required to cancel the divergent hierarchy contributions to the Higgs mass~\cite{Arkani2002}\nocite{Mandl1984}. Of course, if the Higgs particle does not exist, then the hierarchy problem is obviously eliminated. This will be the solution of the hierarchy problem that we have proposed~\cite{Moffat1991,Clayton1991b,Moffat2007f}.

The LHC with a center-of-mass energy, $\sqrt{s}=14$ TeV, should be able to measure the $W_LW_L$ scattering and determine whether the symmetry breaking mechanism is weakly or strongly interacting. These vector bosons are produced in a fermion scattering process, in which the four-fermion interaction is replaced by vector boson exchange in the Lagrangian. For instance, the Lagrangian that describes charged weak current interactions between leptons ($l,~\nu$) and vector bosons can be written in the form
\begin{equation}
{\cal L}_\mathrm{lcc}=\frac{ig}{2\sqrt{2}}\big[W_\mu^+(\bar\nu\gamma^\mu(1-\gamma_5)l)+W_\mu^-(\bar l\gamma^\mu(1-\gamma_5)\nu)\big],
\end{equation}
where $g$ is the $SU_L(2)$ gauge coupling constant. The scattering amplitude is given by, for instance,
\begin{equation}
{\cal M}(\bar\nu_e e\rightarrow W^-\rightarrow\mu\bar\nu_\mu)=i\frac{g^2}{8}\big[\bar{e}\gamma_\mu(1-\gamma_5)\nu_e\big]\big[\bar\nu_\mu\gamma_\nu(1-\gamma_5)\mu\big]\left(\frac{\eta^{\mu\nu}-\frac{p^\mu p^\nu}{m^2_W}}{p^2-m_W^2}\right),
\end{equation}
where $p=p_e+p_{\bar\nu_e}$ is the momentum of the exchanged $W$ with mass $m_W$ and $e,\nu_e,...$ denote the Dirac spinors. At low energies $p^2\ll m_W^2$, we obtain the four-fermion interaction
\begin{equation}
{\cal M}_\mathrm{Fermi}(\bar\nu_e e\rightarrow\mu\bar\nu_\mu)=i\frac{G_F}{\sqrt{2}}[\bar e\gamma_\mu(1-\gamma_5)\nu_e][\bar\nu_\mu\gamma^\mu(1-\gamma_5)\mu],
\end{equation}
where
\begin{equation}
G_F=\frac{g^2}{4\sqrt{2}m_W^2}\simeq 1.166\times 10^{-5}\,{\rm GeV}^{-2}.
\end{equation}
is the Fermi constant.

In Section~\ref{sec:local}, we derive the model, beginning with the massless gauge invariant theory. This follows the derivation of the earlier papers~\cite{Moffat1991,Clayton1991b,Moffat2007f} with additional details of the fundamental non-Abelian $SU(2)$ gauge invariant aspects including a manifestly Becchi, Rouet, Stora and Tyutin (BRST \cite{BRS1976,Tyutin1975}) invariant action and the elaboration of the generating function for the path integral formalism. The regularized massless and gauge invariant EW model is developed in Section~\ref{sec:regularized}. In Section~\ref{sec:symbreak}, we explain the symmetry breaking mechanism that induces masses for vector bosons. This is followed, in Section~\ref{sec:measure}, by a derivation of a symmetry breaking measure in the path integral, leading to the breaking of $SU_L(2)\times U_Y(1)$ symmetry of the massless action and the generation of $W$ and $Z$ masses, retaining a zero mass photon. The predicted masses of the $W$ and $Z$ bosons are proportional to a regularizing energy scale $\Lambda_W$. This energy scale is derived from a self-consistency equation involving the quark and lepton internal loops in the self-energy graphs and is determined to be $\Lambda_W\simeq 542$~GeV. We note that {\it this is not a naive cutoff}. The regularization scheme preserves gauge invariance, unitarity and Poincar\'e invariance in the massless limit, and it does not lead to a conflict with low energy electroweak precision data.

In contrast to the standard EW model with a nonzero Higgs field vacuum expectation value that generates the boson and fermion masses from the classical tree graphs, the Higgsless model acquires boson and fermion masses from the quantum loop graphs and not the {\it massless} classical tree graphs. Thus, the mass generating mechanism in FEW is a purely quantum field theory mechanism associated with a symmetry breaking without a classical scalar field Higgs mechanism.

An important feature of the standard EW model with a Higgs particle is that for $W_LW_L\rightarrow W_LW_L$ scattering, the amplitude and cross section for longitudinal scattering of the $W$ bosons do not violate unitarity. We have demonstrated in a separate article~\cite{Moffat2008c}, that in the finite EW model the $W_LW_L\rightarrow W_LW_L$ scattering does not violate unitarity due to the damping at high energies of the self-energy $W$ boson vertices in the scattering amplitudes. Thus, the finite Higgsless model remains unitary above $\sqrt{s} > 1$ TeV. The unitary $W_LW_L\rightarrow W_LW_L$ amplitudes predicted by the Higgsless model are compared to the amplitudes predicted by the standard EW model for various values of the Higgs mass. These predictions can be used to distinguish the Higgsless model from the standard EW model with Higgs exchange.

In Section~\ref{sec:fermass}, we consider the problem of deriving fermion masses. In the standard EW model, the fermion masses are obtained from a spontaneous symmetry breaking mechanism. The simplest version postulates an $SU_L(2)$ doublet Higgs field $\phi$ and an $SU_L(2)$ invariant Yukawa coupling to the fermions \cite{Burgess2007}:
\begin{equation}
{\cal L}_\mathrm{fermion~masses}=-(f_{mn}\bar L_mP_RE_n\phi+g_{mn}\bar Q_mP_RU_n\tilde\phi+h_{mn}\bar Q_mP_RD_n\phi+\mathrm{h.c.}),
\end{equation}
where $L_m$, $Q_m$ denote lepton and quark doublets, $E_n$, $U_n$, $D_n$ denote charged lepton, up-type quark, and down-type quark singlets, and the indices $n,m=1...3$ run through the three fermion generations. The indices of the Higgs doublet $\phi$ and its conjugate $\tilde\phi$ match against the indices of the left-handed quark and lepton doublets. The abbreviation h.c. stands for hermitian conjugate. The matrices $f_{mn}$, $g_{mn}$, and $h_{mn}$ can be diagonalized to $f_m$, $g_m$, and $h_m$, respectively, and after taking into account the vacuum expectation value $v$ of the Higgs doublet, we arrive at the final form of the fermion mass term:
\begin{equation}
{\cal L}_\mathrm{fermion~masses}=-\frac{1}{\sqrt{2}}v(f_m\bar e_me_m+g_m\bar u_mu_m+h_m\bar d_md_m),
\end{equation}
with the charged lepton, up-type quark, and down-type quark fields represented by the Dirac spinors $e$, $u$, and $d$.

In our FEW model, the fermion masses and the $W$ and $Z$ boson masses are generated by a non-perturbative mass gap equation determined by the fermion\footnote{Or from a finite four-fermion interaction involving the quarks and leptons, which we used in earlier versions of our work \cite{Moffat1991,Clayton1991b}.} and boson self-energy loop graphs \cite{Moffat2007f}, respectively.

\section{The Gauge Invariant Local Theory}
\label{sec:local}

We shall use the metric convention, $\eta_{\mu\nu}={\rm diag}(+1,-1,-1,-1)$, and set $\hbar=c=1$. The theory is based on the local $SU_L(2)\times U_Y(1)$ invariant Lagrangian that includes leptons and quarks (with the color degree of freedom of the strong interaction group $SU_c(3)$) and the boson vector fields that arise from gauging the $SU_L(2)\times U_Y(1)$ symmetry:
\begin{equation}
\label{LocalLag}
L_\mathrm{local}=L_F+L_W+L_B+L_I.
\end{equation}
$L_F$ is the free fermion Lagrangian consisting of massless kinetic terms for each fermion:
\begin{equation}
L_F=\sum_\psi\bar\psi i\slashed\partial\psi=\sum_{q^L}\bar q^Li\slashed\partial q^L+\sum_f\bar\psi^Ri\slashed\partial
\psi^R,
\end{equation}
where the fermion fields have been rewritten as $SU_L(2)$ doublets:
\begin{equation}
q^L\in\left[\begin{pmatrix}\nu^L\\e^L\end{pmatrix},\begin{pmatrix}u^L\\d^L\end{pmatrix}_{r,g,b}\right]
\end{equation}
and U(1)$_Y$ singlets, and we have suppressed the fermion generation indices. We have written $\psi_{L,R}=\frac{1}{2}P_{L,R}\psi$, where $P_{L,R}=\frac{1}{2}(1\mp\gamma_5)$. The Abelian kinetic contribution is given by
\begin{equation}
L_B=-\frac{1}{4}B^{\mu\nu}B_{\mu\nu},
\end{equation}
where
\begin{equation}
B_{\mu\nu}=\partial_\mu B_\nu-\partial_\nu B_\mu.
\end{equation}
The non-Abelian contribution is
\begin{equation}
L_W=-\frac{1}{4}W_{\mu\nu}^aW^{a\mu\nu},
\end{equation}
where
\begin{equation}
W^a_{\mu\nu}=\partial_\mu W_\nu^a-\partial_\nu W_\mu^a-gf^{abc}W_\mu^bW_\nu^c.
\end{equation}

The $SU(2)$ generators satisfy the commutation relations
\begin{equation}
[T^a,T^b]=if^{abc}T^c,~~~~~\mathrm{with}~~~~~T^a=\frac{1}{2}\sigma^a.
\end{equation}
Here, $\sigma^a$ are the Pauli spin matrices and $f^{abc}=\epsilon^{abc}$. The fermion--gauge boson interaction terms are contained in
\begin{equation}
L_I=-gJ^{a\mu}W_\mu^a-g'J_Y^\mu B_\mu,
\end{equation}
where the $SU(2)$ and hypercharge currents are given by
\begin{equation}
J^{a\mu}=\sum_{q^L}\bar{q}^L\gamma^\mu T^aq^L,~~~~~\mathrm{and}~~~~~J_Y^\mu=\sum_\psi\frac{1}{2}Y_\psi\bar\psi\gamma^\mu\psi,
\end{equation}
respectively. The last sum is over all left and right-handed fermion states with hypercharge factors $Y=2(Q-T^3)$:
\begin{align}
Y(q_\mathrm{lepton}^L)=-1,~~~Y(q_\mathrm{quark}^L)=\frac{1}{3},~~~Y(e^R)=-2,\nonumber\\
Y(\nu^R)=0,~~~Y(u^R)=\frac{4}{3},~~~Y(d^R)=\frac{2}{3}.
\end{align}
We also define for notational convenience:
\begin{equation}
\slashed W=\gamma^\mu W_\mu^aT^a.
\end{equation}

The Lagrangian (\ref{LocalLag}) is invariant under the following local gauge transformations (to order $g,g'$):
\begin{align}
\delta W_\mu^a=\partial_\mu\theta^a+gf^{abc}\theta^b W_\mu^c,~~~&~~~\delta B_\mu=\partial_\mu\beta,\nonumber\\
\delta\psi^L=-\left(igT^a\theta^a+ig'\frac{Y_\psi}{2}\beta\right)\psi^L,~~~&~~~\delta\psi^R=-ig'\frac{Y_\psi}{2}
\beta\psi^R,
\label{localtransformations}
\end{align}
giving us an $SU_L(2)\times U_Y(1)$ invariant Lagrangian. Quantization is accomplished via the path integral formalism, which gives the expectation value of operators ${\cal O}[\phi]$ as a sum over all field configurations weighted by the exponential of the classical action:
\begin{equation}
\label{expectationvalue}
\left<T({\cal O}[\phi])\right>\propto\int[D\bar\psi][D\psi][DW][DB]\mu_\mathrm{inv}[\bar\psi,\psi,B,W]{\cal O}[\phi]
\exp\left(i\int d^4xL_\mathrm{local}\right),
\end{equation}
where in the local case the invariant measure $\mu_\mathrm{inv}$ is the trivial one. As it stands, this expression is infinite due to the gauge invariance of the Lagrangian. That is, there is an infinite number of field configurations all related by gauge transformations that contribute equal amounts to any expectation value. To remedy this situation, we introduce gauge fixing terms:
\begin{equation}
L_\mathrm{gf}=-\frac{1}{2\xi}(\partial_\mu B^\mu)^2-\frac{1}{2\xi}(\partial_\mu W^{a\mu})^2.
\end{equation}
Here, the choice of the gauge parameter $\xi$ could be different for each gauge field, but for simplicity we have chosen the same gauge condition for the Abelian and non-Abelian gauge bosons. As we require gauge invariant results, this constraint should not cause any physical prediction to pick up a dependence on the gauge parameter $\xi$. We ensure this by introducing auxiliary ghost fields into the theory~\cite{Itzykson1980,Zee2003,Peskin1995}.

If we consider a general gauge condition: ${\cal F}_a(A)^2=0$ ($A$ stands here for either gauge field), then a particular choice of ${\cal F}$ should not affect the expectation value (\ref{expectationvalue}). We can guarantee this by introducing the Jacobian determinant for the transformation into the path integral:
\begin{equation}
\left<T({\cal O}[\phi])\right>\propto\int[DA]\det[{\cal M}]{\cal O}[\phi]\exp\left(i\int d^4x[L_\mathrm{local}+L_\mathrm{gf}]\right).
\end{equation}
The operator in the determinant is given by
\begin{equation}
{\cal M}_{ab}(x,y)=\frac{\delta}{\delta\theta^b(y)}\left[\frac{\delta{\cal F}_a}{\delta A^{c\mu}(x)}\delta A^{c\mu}(x)\right]_{\theta=0}.
\end{equation}
In the case of the Abelian fields
\begin{equation}
{\cal M}_Y(x,y)=\Box\delta^4(x-y),
\end{equation}
while for the non-Abelian fields we get
\begin{equation}
{\cal M}_{ab}(x,y)=\left[\Box\delta_{ab}+gf_{abc}\partial^\mu W_\mu^c\right]\delta^4(x-y),
\end{equation}
where $\Box=\partial_\mu\partial^\mu$.

If the determinant is left in the perturbative expansion, it results in nonlocal interaction terms, but by using a Grassmann algebra, one can rewrite the determinant in terms of auxiliary ghost fields
\begin{equation}
\det[i{\cal M}]=\int[D\bar\eta][D\eta]\exp(-i\bar\eta^a{\cal M}^{ab}\eta^b)
\end{equation}
which results in another piece in the effective Lagrangian
\begin{equation}
L_\mathrm{ghost}=\bar\eta\Box\eta+\bar c^a\Box c^a-gf^{abc}\bar c^a\partial_\mu(W^{b\mu}c^c).
\end{equation}
We then have the final form of the path integral in the quantized theory:
\begin{equation}
\left<T({\cal O}[\phi])\right>\propto\int[D\bar\psi][D\psi][DW][DB][D\bar\eta][D\eta][D\bar c][Dc]{\cal O}[\phi]\exp(iS_\mathrm{eff}),
\end{equation}
where the effective action is given by
\begin{equation}
\label{effaction}
S_\mathrm{eff}=\int d^4x(L_\mathrm{local}+L_\mathrm{gf}+L_\mathrm{ghost})=\int d^4x(L_F+L_I).
\end{equation}
Here, we have separated the Lagrangian into quadratic and interaction pieces.

The action (\ref{effaction}) is not invariant under the gauge transformation given earlier in equation~(\ref{localtransformations}). Successful quantization of the theory implies invariance under an extended set of BRST transformations, generated by replacing the infinitesimal fields $\theta^a$ and $\beta$ by
\begin{equation}
\theta^a\rightarrow-c^a\lambda\xi,~~~~~\beta\rightarrow-\eta\lambda_0\xi,
\end{equation}
where $\lambda$ and $\lambda_0$ are infinitesimal Grassmann constants. This generates the transformations
\begin{equation}
\delta W_\mu^a=\lambda\xi \partial_\mu c^a+g\lambda\xi f^{abc}c^bW_\mu^c,~~~~~\delta B_\mu=\lambda_0\xi\partial_\mu\eta,
$$ $$
\delta\psi^L=\left(-ig\lambda\xi T^ac^a-ig'\xi\lambda_0\frac{Y_\psi}{2}\eta\right)\psi^L,~~~~~\delta\psi^R=-ig'
\xi\lambda_0\frac{Y_\psi}{2}\eta\psi^R,
\end{equation}
which leave $L_\mathrm{local}$ invariant, and also leave $S_\mathrm{eff}$ invariant provided the ghost fields transform as
\begin{align}
\delta c^a=-\frac{\xi}{2}\lambda gf^{abc}c^bc^c,~~~~~&~~~~~\delta\bar c^a=\lambda \partial_\mu W^{a\mu},\nonumber\\
\delta\eta=0,~~~~~&~~~~~\delta\bar\eta=-\lambda_0 \partial_\mu B^\mu.
\end{align}
We now have a correctly quantized theory and we need to generate the perturbative expansion.

Using $\phi$ to denote any field and $J$ to denote a generic source term, we introduce the solution of the classical free-field equation (quadratic terms) in the presence of a source term:
\begin{equation}
\Delta^{-1}(x)\phi_c(x)=-J(x),
\end{equation}
which gives as a solution to the classical equation
\begin{equation}
\phi_c(x)=\int i\Delta(x-y)iJ(y)d^4y.
\end{equation}
We can then write the generating functional for connected diagrams as
\begin{equation}
W[J]=\ln(Z[J])=\ln\left[\int[D\phi]\exp\left(i\int dx(L_q[\phi]+L_I[\phi]+J\phi)\right)\right],
\end{equation}
where $L_q$ denotes the quadratic parts of the Lagrangian. Then, we have
\begin{equation}
Z[J]=\exp\left(i\int L_I\left[\frac{1}{i}\frac{\delta}{\delta J(x)}\right]d^4x\right)Z_0[J],
\end{equation}
where
\begin{align}
Z_0[J]&=\int[D\phi]\exp\left(i\int dx(L_q[\phi]+J\phi)\right)\nonumber\\
&\propto\exp\left(\frac{1}{2}\int d^4xd^4y(iJ(x)i
\Delta(x-y)iJ(y))\right).
\end{align}
The connected Green's functions and thereby the usual Feynman rules are then generated by
\begin{equation}
{\cal G}(x_1,...,x_n)=\left<0|T[\phi(x_1)...\phi(x_n)]|0\right>=i^n\frac{\delta^nW[J]}{\delta J(x_1)...\delta J(x_n)}.
\end{equation}
The Green's functions give the momentum space two-point propagators of the theory:
\begin{equation}
iS=\frac{-i}{\slashed p+i\epsilon}=\frac{-i\slashed p}{p^2+i\epsilon},\quad i\Delta=\frac{-i}{p^2+i\epsilon},
\end{equation}
and
\begin{equation}
iD_{\mu\nu}=\frac{-i}{p^2+i\epsilon}\left(\eta_{\mu\nu}+(\xi-1)\frac{p_\mu p_\nu}{p^2}\right),\quad i\Delta^{ab}=i\Delta\delta^{ab},~~~~~iD_{\mu\nu}^{ab}=iD_{\mu\nu}\delta^{ab}.
\end{equation}
\begin{figure}[t]
\begin{center}
\begin{minipage}{0.59\linewidth}
\begin{minipage}{0.39\linewidth}
\begin{fmfgraph*}(80,40)
\fmfleftn{i}{1}\fmfrightn{o}{2}
\fmf{wiggly,label=$p_1$,lab.side=left}{i1,v1}
\fmf{dots_arrow,label=$p_2$,lab.side=left}{v1,o2}
\fmf{dots_arrow,label=$p_3$,lab.side=left}{o1,v1}
\fmflabel{${}_\alpha^a$}{i1}
\fmflabel{${}^b$}{o2}
\fmflabel{${}^c$}{o1}
\end{fmfgraph*}
\end{minipage}
\begin{minipage}{0.49\linewidth}
\begin{flushleft}
$\lefteqn{igI_\alpha^{abc}(p_1,p_2,p_3)}$
\end{flushleft}
\end{minipage}
\\\vskip 3em
\begin{minipage}{0.39\linewidth}
\begin{fmfgraph*}(80,40)
\fmfleftn{i}{1}\fmfrightn{o}{2}
\fmf{wiggly,label=$p_1$,lab.side=left}{i1,v1}
\fmf{wiggly,label=$p_3$,lab.side=left}{v1,o2}
\fmf{wiggly,label=$p_2$,lab.side=left}{o1,v1}
\fmflabel{${}_\alpha^a$}{i1}
\fmflabel{${}_\beta^b$}{o1}
\fmflabel{${}_\gamma^c$}{o2}
\end{fmfgraph*}
\end{minipage}
\begin{minipage}{0.49\linewidth}
\begin{flushleft}
$igI_{\alpha\beta\gamma}^{abc}(p_1,p_2,p_3)$
\end{flushleft}
\end{minipage}
\\\vskip 3em
\begin{minipage}{0.39\linewidth}
\begin{fmfgraph*}(80,40)
\fmfleftn{i}{2}\fmfrightn{o}{2}
\fmf{wiggly,label=$p_2$,lab.side=right}{i1,v1}
\fmf{wiggly,label=$p_1$,lab.side=left}{i2,v1}
\fmf{wiggly,label=$p_3$,lab.side=left}{o1,v1}
\fmf{wiggly,label=$p_4$,lab.side=right}{o2,v1}
\fmflabel{${}_\alpha^a$}{i2}
\fmflabel{${}_\beta^b$}{i1}
\fmflabel{${}_\gamma^c$}{o1}
\fmflabel{${}_\delta^d$}{o2}
\end{fmfgraph*}
\end{minipage}
\begin{minipage}{0.49\linewidth}
\begin{flushleft}
$ig^2I_{\alpha\beta\gamma\delta}^{abcd}(p_1,p_2,p_3,p_4)$
\end{flushleft}
\end{minipage}
\end{minipage}
\begin{minipage}{0.39\linewidth}
\begin{minipage}{0.49\linewidth}
\begin{fmfgraph*}(80,40)
\fmfleftn{i}{1}\fmfrightn{o}{2}
\fmf{wiggly,label=$p$,lab.side=left}{i1,v1}
\fmf{fermion}{v1,o1}
\fmf{fermion}{o2,v1}
\fmflabel{${}_\alpha^a$}{i1}
\fmflabel{$q_1$}{o2}
\fmflabel{$q_2$}{o1}
\end{fmfgraph*}
\end{minipage}
\begin{minipage}{0.49\linewidth}
\begin{flushleft}
$igF^{a\alpha}$
\end{flushleft}
\end{minipage}
\\\vskip 3em
\begin{minipage}{0.49\linewidth}
\begin{fmfgraph*}(80,40)
\fmfleftn{i}{1}\fmfrightn{o}{2}
\fmf{wiggly,label=$p$,lab.side=left}{i1,v1}
\fmf{fermion}{v1,o1}
\fmf{fermion}{o2,v1}
\fmflabel{${}_\alpha$}{i1}
\fmflabel{$q_1$}{o2}
\fmflabel{$q_2$}{o1}
\end{fmfgraph*}
\end{minipage}
\begin{minipage}{0.49\linewidth}
\begin{flushleft}
$ig'F^\alpha$
\end{flushleft}
\end{minipage}
\end{minipage}
\end{center}
\caption{The vertex rules of the gauge invariant local theory.}
\label{fig:vertices}
\end{figure}
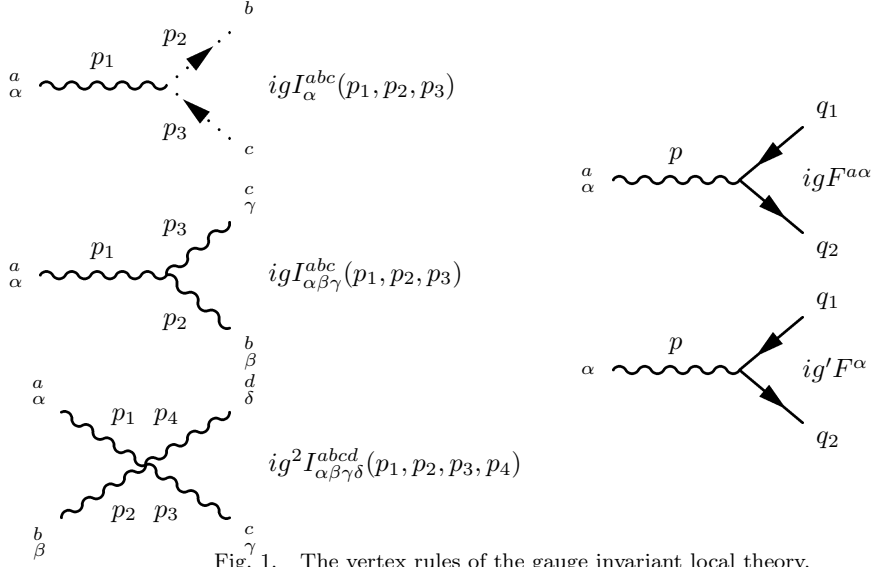
The interaction vertices corresponding to Figure~\ref{fig:vertices} are given by
\begin{align}
I_{\alpha}^{abc}(p_1,p_2,p_3)=&-if^{abc}p_{2\alpha}\\
I_{\alpha\beta\gamma}^{abc}(p_1,p_2,p_3)=&-if^{abc}[(p_2-p_1)_\gamma g_{\alpha\beta}+(p_3-p_2)_\alpha g_{\beta\gamma}+(p_1-p_3)_\beta g_{\gamma\alpha}]
\end{align}
\begin{align}
I_{\alpha\beta\gamma\delta}^{abcd}(p_1,p_2,p_3,p_4)=&-[f^{abc}f^{cde}(g_{\alpha\gamma}g_{\beta\delta}-g_{\gamma\beta}g_{\delta\alpha})\nonumber\\
&~~~+f^{ace}f^{dbe}(g_{\alpha\delta}g_{\gamma\beta}-g_{\delta\gamma}g_{\alpha\beta})\nonumber\\
&~~~+f^{abc}f^{cde}(g_{\alpha\beta}g_{\delta\gamma}-g_{\beta\delta}g_{\gamma\alpha})]
\end{align}
\begin{align}
F^{a\alpha}=&T^a\gamma^\alpha P_LF^\alpha=-\gamma^\alpha\left(\frac{Y_L}{2}P_L+\frac{Y_R}{2}P_R\right)=-\gamma^\alpha(Q-T^3P_L).
\end{align}
Since we want to have a gauge invariant perturbation scheme, we also require that the generating functional is invariant under the BRST transformation. This generates the usual Ward-Takahashi identities that the irreducible vertex functions and dressed propagators must satisfy.

Finally, we look at diagonalizing the charged sector and mixing in the neutral boson sector. If we write
\begin{equation}
W^\pm=\frac{1}{\sqrt{2}}(W^1\mp iW^2)
\end{equation}
as the physical charged vector boson fields, then we get the fermion interaction terms:
\begin{equation}
-\frac{g}{\sqrt{2}}(J_\mu^+W^{+\mu}+J_\mu^-W^{-\mu}),
\end{equation}
where the charged current is given by
\begin{equation}
J_\mu^\pm=J_{1\mu}^\pm\pm iJ_{2\mu}^\pm=\sum_{q_L}\bar q^L\gamma_\mu T^\pm q^L~~~~~{\rm implying}~~~~~J_\mu^+=\sum_{q_L}(\bar\nu^L\gamma_\mu e^L+\bar u^L\gamma_\mu d^L).
\end{equation}
In the neutral sector, we can mix the fields in the usual way:
\begin{equation}
Z_\mu=c_wW_\mu^3-s_wB_\mu~~~\mathrm{and}~~~A_\mu=c_wB_\mu+s_wW_\mu^3,
\label{eq:2.35}
\end{equation}
where $s_w=\sin\theta_w$ and $c_w=\cos\theta_w$ with $\theta_w$ denoting the weak mixing (Weinberg) angle. We define the usual trigonometric relations
\begin{equation}
s^2_w=\frac{g'^2}{g^2+g'^2}~~~\mathrm{and}~~~c^2_w=\frac{g^2}{g^2+g'^2}.
\end{equation}
The neutral current fermion interaction terms now look like:
\begin{equation}
-gJ^{3\mu}W_\mu^3-g'J_Y^\mu B_\mu=-(gs_wJ^{3\mu}+g'c_wJ_Y^\mu)A_\mu-(gc_wJ^{3\mu}-g's_wJ_Y^\mu)Z_\mu.
\end{equation}
If we identify the resulting $A_\mu$ field with the photon, then we have the unification condition:
\begin{equation}
e=gs_w=g'c_w
\end{equation}
and the electromagnetic current is
\begin{equation}
J_\mathrm{em}^\mu=J^{3\mu}+J_Y^\mu,
\end{equation}
where $e$ is the charge of the proton. Note that the coupling now looks like $(Q-T^3)+T^3=Q$ and we only get coupling of the photon to charged fermions at tree level. We can then identify the neutral current:
\begin{equation}
J_\mathrm{NC}^\mu=J^{3\mu}-s_wJ_\mathrm{em}^\mu,
\end{equation}
and write the fermion-boson interaction terms as
\begin{equation}
L_I=-\frac{g}{\sqrt{2}}(J_\mu^+W^{+\mu}+J_\mu^-W^{-\mu})-gs_wJ_\mathrm{em}^\mu A_\mu-\frac{g}{c_w}J_\mathrm{NC}^\mu Z_\mu.
\end{equation}
This, along with the suitably rewritten boson interaction terms, gives the usual vertices of the local point theory.

\section{The Gauge Invariant Regularized Theory}
\label{sec:regularized}

To write the theory in its finite, nonlocal form, we follow the method outlined in refs.~\cite{Moffat1989,Moffat1990,Evens1991,Moffat1991b,Hand1991,Kleppe1991,Kleppe1992,Cornish1992b,Cornish1992c,Cornish1992,Kleppe1993,Clayton1994,Paris1995,Paris1996,Saini1997,Efimov1967,Efimov1968,Efimov1966,Efimov1967b}.

The key observation in these cited works, most notably \cite{Evens1991}, is that when the vertices of a theory contain nonlocal factors (as they do in string theory, for instance), this causes loops to converge in Euclidean space and {\em any} otherwise local Lagrangian gives an ultraviolet-finite theory. Accordingly, to regularize the fields we write the non-local (smeared) fields as a convolution of the local fields with a function whose momentum space Fourier transform is an {\it entire function}, which is complex differentiable everywhere in the complex plane, and thus it does not introduce unphysical poles into the propagators of the theory. This function can be related to a Lorentz invariant operator distribution as~\cite{Moffat1990,Evens1991}:
\begin{equation}
\Phi(x)=\int d^4yG(x-y)\phi(y)=G\biggl(\frac{\Box}{\Lambda_W^2}\biggr)\phi(x),
\end{equation}
where $\Lambda_W$ denotes a non-local electroweak energy scale. We make a choice of a specific smearing operator:
\begin{equation}
G\left(\frac{\Box}{\Lambda_W^2}\right)\equiv{\cal E}_m=\exp\left(-\frac{\Box+m^2}{2\Lambda_W^2}\right).
\end{equation}

This procedure destroys (local) gauge invariance \cite{Evens1991}. We restore gauge invariance and ensure that the tree graphs remain local and point-like by introducing additional interaction terms into the Lagrangian, enforcing decoupling of unphysical degrees of freedom. This procedure can be repeated, order by order, as shown in \cite{Evens1991}\footnote{In \cite{Evens1991}, the authors concern themselves primarily with QED; however, in the last section of the paper, the non-Abelian case is discussed and proof is offered that there is at least one solution to all orders.}. Current conservation and the Ward identities for the nonlocal symmetry follow by changing variables in the usual manner.

Given a gauge invariant classical theory, quantization through the path integral formalism can proceed, but problems may arise due to the functional measures $[D\psi]$, $[D\bar\psi]$, $[DW]$, and $[DB]$. Because the transformation rule for the Abelian field $B$ is unchanged, we only need to consider the behavior of $[DW]$, $[D\psi]$ and $[D\bar\psi]$ under a gauge transformation. Invariance can be restored by finding an ``acceptable'' measure factor that generates additional interactions that restore gauge invariance. Finding an acceptable measure is non-trivial, but for QED, such a measure has been proven to exist \cite{Evens1991}. We address a necessary condition of gauge invariance, which is the vanishing of fermionic and $W$ masses. Operationally, one may aim to satisfy this condition by looking at the self-energies to second order and demanding transversality of the vacuum polarization tensor. This is the route that we follow below.

A genuine anomaly will show up here as the non-existence of a measure factor, but as we shall see, we are safe at this order. We observe that this theory is only rigorously defined in Euclidean space, but since it has been shown that an analytic continuation to Minkowski space always exists via Efimov's regulator \cite{Efimov1967,Efimov1968,Efimov1966,Efimov1967b}, we will work in Minkowski space, only referring to Euclidean space to ensure the convergence of the loop integrals.

When we regularize a theory that is initially massless, all fields are smeared with ${\cal E}_0$. We now write the initial Lagrangian in non-local form:
\begin{equation}
L_\mathrm{reg}=L[\phi]_F+{\cal L}[\Phi]_I,
\end{equation}
where ${\cal L}[\Phi]_I$ indicates smearing of the interacting fields.

An essential feature of the regularized, non-local field theory is the requirement that the classical tree graph theory remain local, giving us a well defined classical limit in the gauge invariant case. Before we proceed, we make use of a field redefinition. We first note that we must alter the quantized form of the theory by generalizing the path integral~\cite{Moffat2007f,Evens1991}:
\begin{equation}
\left<T^*({\cal O}[\Phi])\right>\propto\int[D\bar\psi][d\psi][DW][DB][D\bar\eta][D\eta][D\bar c][Dc]{\cal O}
[\Phi]\exp(iS_0[\phi]+iS_I[\Phi]),
\end{equation}
where we are now dealing with expectation values of operators that are functionals of the smeared fields $\Phi$.

To generate a perturbation scheme in the non-local operators, we write the generating functional as
\begin{equation}
W[{\cal J}]=\ln(Z[{\cal J}])=\ln\left(\int[D\phi]\exp\left(i\int dx\{L_F[\phi]+{\cal L}_I[\Phi]
+{\cal J}(x)\Phi(x)\}\right)\right),
\end{equation}
where the source term ${\cal J}$ is now non-local. We note that in momentum space, the smeared fields are related one-to-one to the local fields:
\begin{equation}
\Phi(p)=G(p^2)\phi(p)=\exp\left(\frac{p^2-m^2}{2\Lambda_W^2}\right)\phi(p),
\end{equation}
so we can take $\phi\rightarrow G^{-1}\phi$ to give
\begin{equation}
\left<T^*({\cal O}[\Phi])\right>\propto\int[D\bar\psi][D\psi][DW][DB][D\bar\eta][D\eta][D\bar c][Dc]
{\cal O}[\phi]\exp\left(iS_0[G^{-1}\phi]+iS_I[\phi]\right),
\end{equation}
leaving the interaction vertices identical to those of the local theory and causing the propagators to pick up a factor of $G^2$. Although this alters the Feynman rules of the theory, it does not alter any physical quantities generated by them.

Now we could continue as we did in the local case to quantize the theory by introducing gauge fixing terms, but we would find that the now non-local gauge symmetry of ${\cal L}_\mathrm{non-local}$ would make finding the Faddeev-Popov determinant difficult and add higher order ghost interactions to the theory. Instead, we will begin with the BRST invariant local Lagrangian directly and derive the nonlocal theory. We begin with the quadratic Lagrangian:
\begin{align}
L_F=&\sum_\psi\bar\psi\frac{i\slashed\partial}{G^2}\psi-\frac{1}{4}(\partial^\mu B^\nu-\partial^\nu B^\mu)\frac{1}{G^2}(\partial_\mu B_\nu-\partial_\nu B_\mu)\\
&-\frac{1}{4}(\partial^\mu W^{a\nu}-\partial^\nu W^{a\mu})\frac{1}{G^2}(\partial_\mu W_{\nu}^a-\partial_\nu W_\mu^a)\nonumber\\
&+\bar c^a\frac{\Box}{G^2}c^a+\bar\eta\frac{\Box}{G^2}\eta-\frac{1}{2\xi}\partial_\mu B^\mu\frac{1}
{G^2}\partial_\nu B^\nu
-\frac{1}{2\xi}\partial_\mu W^{a\mu}\frac{1}{G^2}\partial_\nu W^{a\nu}\nonumber\\
=&\sum_\psi\bar\psi\frac{S^{-1}}{G^2}\psi-\frac{1}{2}B^\mu\frac{(D_{\mu\nu})^{-1}}{G^2}B^\nu-W^{a\mu}
\frac{(D_{\mu\nu}^{ab})^{-1}}{G^2}W^{b\nu}+\bar c^a\frac{(\Delta^{ab})^{-1}}{G^2}c^b+\bar\eta\frac{\Delta^{-1}}{G^2}\eta,\nonumber
\end{align}
which generates propagators that are $G^2$ multiplied by those given in the local theory. For convenience later on, we will define another set of propagators as $(1-G^2)$ times the local ones:
\begin{equation}
i\bar\Delta=(1-G^2)i\Delta,~~~~~\mathrm{etc.},
\end{equation}
so that the sum of the tree propagators with these give the causal propagators of point theory. This is useful when calculating tree graphs, since one can merely replace the smeared propagator with the barred one in the amplitude, and then add the appropriate term to the interaction Lagrangian. This procedure guarantees that all calculated tree graphs are local and point-like to all orders of perturbation theory~\cite{Evens1991}.

Along with the interaction terms of the local theory that now look identical after having made the field redefinition, we have to second order in coupling additional terms coming from fixing the tree graphs in Figure~\ref{fig:fixgraphs}:
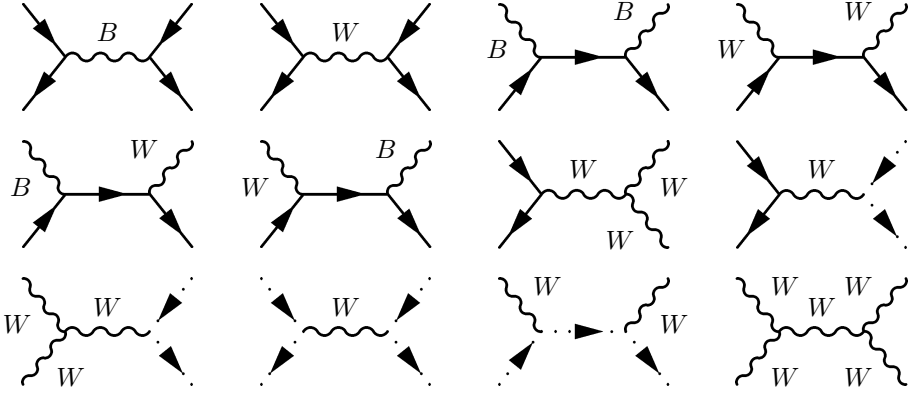
\begin{figure}
\begin{minipage}{0.24\linewidth}
\begin{fmfgraph*}(80,40)
\fmfleftn{i}{2}\fmfrightn{o}{2}
\fmf{fermion}{v1,i1}
\fmf{fermion}{i2,v1}
\fmf{wiggly,label=$B$,lab.side=left}{v1,v2}
\fmf{fermion}{v2,o1}
\fmf{fermion}{o2,v2}
\end{fmfgraph*}
\end{minipage}
\begin{minipage}{0.24\linewidth}
\begin{fmfgraph*}(80,40)
\fmfleftn{i}{2}\fmfrightn{o}{2}
\fmf{fermion}{v1,i1}
\fmf{fermion}{i2,v1}
\fmf{wiggly,label=$W$,lab.side=left}{v1,v2}
\fmf{fermion}{v2,o1}
\fmf{fermion}{o2,v2}
\end{fmfgraph*}
\end{minipage}
\begin{minipage}{0.24\linewidth}
\begin{fmfgraph*}(80,40)
\fmfleftn{i}{2}\fmfrightn{o}{2}
\fmf{wiggly,label=$B$}{v1,i2}
\fmf{fermion}{i1,v1}
\fmf{fermion}{v1,v2}
\fmf{fermion}{v2,o1}
\fmf{wiggly,label=$B$}{o2,v2}
\end{fmfgraph*}
\end{minipage}
\begin{minipage}{0.24\linewidth}
\begin{fmfgraph*}(80,40)
\fmfleftn{i}{2}\fmfrightn{o}{2}
\fmf{wiggly,label=$W$}{v1,i2}
\fmf{fermion}{i1,v1}
\fmf{fermion}{v1,v2}
\fmf{fermion}{v2,o1}
\fmf{wiggly,label=$W$}{o2,v2}
\end{fmfgraph*}
\end{minipage}
\\\vskip 1em
\begin{minipage}{0.24\linewidth}
\begin{fmfgraph*}(80,40)
\fmfleftn{i}{2}\fmfrightn{o}{2}
\fmf{wiggly,label=$B$}{v1,i2}
\fmf{fermion}{i1,v1}
\fmf{fermion}{v1,v2}
\fmf{fermion}{v2,o1}
\fmf{wiggly,label=$W$}{o2,v2}
\end{fmfgraph*}
\end{minipage}
\begin{minipage}{0.24\linewidth}
\begin{fmfgraph*}(80,40)
\fmfleftn{i}{2}\fmfrightn{o}{2}
\fmf{wiggly,label=$W$}{v1,i2}
\fmf{fermion}{i1,v1}
\fmf{fermion}{v1,v2}
\fmf{fermion}{v2,o1}
\fmf{wiggly,label=$B$}{o2,v2}
\end{fmfgraph*}
\end{minipage}
\begin{minipage}{0.24\linewidth}
\begin{fmfgraph*}(80,40)
\fmfleftn{i}{2}\fmfrightn{o}{2}
\fmf{fermion}{v1,i1}
\fmf{fermion}{i2,v1}
\fmf{wiggly,label=$W$,lab.side=left}{v1,v2}
\fmf{wiggly,label=$W$,lab.side=left}{o1,v2}
\fmf{wiggly,label=$W$,lab.side=left}{o2,v2}
\end{fmfgraph*}
\end{minipage}
\begin{minipage}{0.24\linewidth}
\begin{fmfgraph*}(80,40)
\fmfleftn{i}{2}\fmfrightn{o}{2}
\fmf{fermion}{v1,i1}
\fmf{fermion}{i2,v1}
\fmf{wiggly,label=$W$,lab.side=left}{v1,v2}
\fmf{dots_arrow}{v2,o1}
\fmf{dots_arrow}{o2,v2}
\end{fmfgraph*}
\end{minipage}
\\\vskip 1em
\begin{minipage}{0.24\linewidth}
\begin{fmfgraph*}(80,40)
\fmfleftn{i}{2}\fmfrightn{o}{2}
\fmf{wiggly,label=$W$,lab.side=left}{v1,i1}
\fmf{wiggly,label=$W$,lab.side=left}{v1,i2}
\fmf{wiggly,label=$W$,lab.side=left}{v1,v2}
\fmf{dots_arrow}{v2,o1}
\fmf{dots_arrow}{o2,v2}
\end{fmfgraph*}
\end{minipage}
\begin{minipage}{0.24\linewidth}
\begin{fmfgraph*}(80,40)
\fmfleftn{i}{2}\fmfrightn{o}{2}
\fmf{dots_arrow}{v1,i1}
\fmf{dots_arrow}{i2,v1}
\fmf{wiggly,label=$W$,lab.side=left}{v1,v2}
\fmf{dots_arrow}{v2,o1}
\fmf{dots_arrow}{o2,v2}
\end{fmfgraph*}
\end{minipage}
\begin{minipage}{0.24\linewidth}
\begin{fmfgraph*}(80,40)
\fmfleftn{i}{2}\fmfrightn{o}{2}
\fmf{dots_arrow}{i1,v1}
\fmf{wiggly,label=$W$,lab.side=left}{i2,v1}
\fmf{dots_arrow}{v1,v2}
\fmf{dots_arrow}{v2,o1}
\fmf{wiggly,label=$W$,lab.side=left}{o2,v2}
\end{fmfgraph*}
\end{minipage}
\begin{minipage}{0.24\linewidth}
\begin{fmfgraph*}(80,40)
\fmfleftn{i}{2}\fmfrightn{o}{2}
\fmf{wiggly,label=$W$}{i1,v1}
\fmf{wiggly,label=$W$}{i2,v1}
\fmf{wiggly,label=$W$}{v2,v1}
\fmf{wiggly,label=$W$,lab.side=right}{v2,o1}
\fmf{wiggly,label=$W$,lab.side=left}{v2,o2}
\end{fmfgraph*}
\end{minipage}
\caption{Tree graphs fixed by the nonlocal theory.}
\label{fig:fixgraphs}
\end{figure}
\begin{align}
{\cal L}_I=&-\frac{1}{2}g'^2J_Y^\mu\bar D_{\mu\nu}J_Y^\mu-\frac{1}{2}g^2J^{a\mu}\bar D_{\mu\nu}^{ab}J^{b\nu}-g'^2
\sum_\psi\left(\frac{Y_\psi}{2}\right)^2\bar\psi\slashed B\bar S\slashed B\psi\nonumber\\
&-g^2\sum_{q^L}\bar q^L\slashed W\bar S\slashed Wq^L-gg'\sum_{q^L}\frac{Y_q}{2}\bar q^L\slashed W\bar S\slashed Bq^L-gg'\sum_{q^L}\frac{Y_q}{2}\bar\psi^L\slashed B\bar S\slashed W\psi^L\nonumber\\
&-ig^2J^{a\mu}\bar D_{\mu\nu}^{ab}C^{b\nu}-ig^2f^{acd}J^{b\mu}\bar D_{\mu\nu}^{ab}\partial^\nu\bar c^cc^d+g^2f^{acd}C^{b\mu}\bar D_{\mu\nu}^{ab}\partial^\nu\bar c^cc^d\nonumber\\
&+\frac{1}{2}g^2f^{abc}f^{def}\partial^\mu\bar c^bc^c\bar D_{\mu\nu}^{ad}\partial^\nu\bar c^ec^f-g^2f^{abc}f^{def}\partial^\mu\bar c^bW_\mu^c\bar\Delta^{ad}\partial^\nu(W_\nu^ec^f)\nonumber\\
&+\frac{1}{2}g^2C^{a\mu}\bar D_{\mu\nu}^{ab}C^{b\nu},
\end{align}
where
\begin{equation}
C_\mu^a=f^{abc}(2W_\nu^c\partial^\nu W^b{}_\mu-W_\nu^c\partial_\mu W^{b\nu}-W_\mu^c\partial_\nu W^{b\nu}).
\end{equation}

One can then show that to second order in coupling, ${\cal L}_\mathrm{non-local}$ is invariant under the following non-linear gauge transformations, which can be verified explicitly as being nilpotent by simple algebra:
\begin{align}
\delta W_\mu^a=&\lambda\xi \partial_\mu c^a+g\lambda\xi f^{abc}c^bW_\mu^c-g^2\xi\lambda f^{abc}G^2\big[c^b\bar D_{\mu\nu}C^{c\nu}+c^b\bar D_{\mu\nu}J^{c\nu}\big]\nonumber\\
&-g^2\xi\lambda f^{abc}f^{cde}G^2\big[c^b\bar D_{\mu\nu}\partial^\nu\bar c^dc^e+W^b_\mu\bar\Delta\partial_\nu(W^{d\nu}c^e)\big],
\nonumber\\
\delta B_\mu=&\lambda_0\xi\partial_\mu\eta,
\nonumber\\
\delta\psi^L=&G^2\left[-ig\xi\lambda T^ac^a-ig'\frac{Y_\psi}{2}\xi\lambda_0\eta+ig^2\xi\lambda T^ac^a\bar S\slashed W\right.
\nonumber\\
&\left.+igg'\frac{Y_\psi}{2}\xi\lambda T^ac^a\bar S\slashed B+igg'\frac{Y_\psi}{2}\xi\lambda_0\eta\bar S\slashed W+ig'^2
\left(\frac{Y_\psi}{2}\right)^2\xi\lambda_0\eta\bar S\slashed B\right]\psi^L,
\nonumber\\
\delta\psi^R=&G^2\left[-ig'\xi\lambda_0\frac{Y_\psi}{2}\eta+ig'^2\left(\frac{Y_\psi}{2}\right)^2\xi\lambda_0\eta\bar S
\slashed B\right]\psi^R,
\nonumber\\
\delta c^a=&-\frac{\xi}{2}\lambda gf^{abc}G^2c^bc^c-\xi\lambda g^2f^{abc}f^{cde}G^2c^b\bar\Delta\partial^\mu(W_\mu^dc^e),~~~~~\delta\bar c^a=\lambda\partial_\mu W^{a\mu},
\nonumber\\
\delta\eta=&0,~~~~~\delta\bar\eta=-\lambda_0 \partial_\mu B^\mu.
\end{align}

To prove that the theory is BRST invariant beyond the second order, a different approach is required. Kleppe and Woodard \cite{Kleppe1992} demonstrated that the non-local smearing operator preserves the continuous symmetries of the local action. This proof is directly applicable in the present case as the non-local smearing operator is identical to that used by Kleppe and Woodard. Thus $S_\mathrm{non-local}$ has a BRST invariance, assuring us that we have a correctly quantized theory to all orders. However, there remains the question of whether the entire path integral is invariant. We do not yet have an invariant measure (i.e., $[D\Psi]$ is not invariant under this extended gauge transformation). Indeed, the existence of an invariant measure is not automatically guaranteed; for instance, it has been shown that the invariant measure factor does not exist for the chiral Schwinger model \cite{Hand1992}. Therefore, we must attempt to find the appropriate measure factor by explicit construction.

As noted, we will derive the measure by requiring that the theory remain invariant in the loop expansion. This is equivalent, at second order, to ensuring that nothing picks up a mass term at one loop. We work in the Feynman gauge ($\xi=1$), for it is much simpler operationally to work with, but it should be kept in mind that unphysical degrees of freedom will occur. The simplest self-energy is that of the ghost (see Figure~\ref{fig:ghost}) in Euclidean momentum space $p_E$:
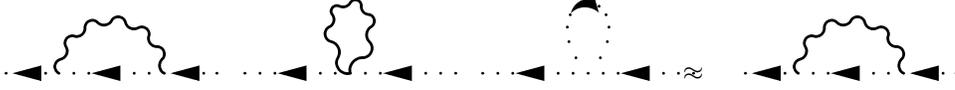
\begin{figure}
\begin{minipage}{0.24\linewidth}
\begin{fmfgraph*}(80,35)
\fmfleftn{i}{1}\fmfrightn{o}{1}
\fmf{dots_arrow}{o1,v2}
\fmf{dots_arrow}{v1,i1}
\fmf{dots_arrow,tension=0.5}{v2,v1}
\fmf{wiggly,left,tension=0}{v1,v2}
\end{fmfgraph*}
\end{minipage}
\begin{minipage}{0.24\linewidth}
\begin{fmfgraph*}(80,35)
\fmfleftn{i}{1}\fmfrightn{o}{1}
\fmf{dots_arrow}{o1,v1}
\fmf{wiggly}{v1,v1}
\fmf{dots_arrow}{v1,i1}
\end{fmfgraph*}
\end{minipage}
\begin{minipage}{0.2\linewidth}
\begin{fmfgraph*}(80,35)
\fmfleftn{i}{1}\fmfrightn{o}{1}
\fmf{dots_arrow}{o1,v1}
\fmf{dots_arrow}{v1,i1}
\fmf{dots_arrow}{v1,v1}
\end{fmfgraph*}
\end{minipage}
$\approx~~~$
\begin{minipage}{0.2\linewidth}
\begin{fmfgraph*}(80,35)
\fmfleftn{i}{1}\fmfrightn{o}{1}
\fmf{dots_arrow}{o1,v2}
\fmf{dots_arrow}{v1,i1}
\fmf{dots_arrow,tension=0.5}{v2,v1}
\fmf{wiggly,left,tension=0}{v1,v2}
\end{fmfgraph*}
\end{minipage}
\vskip -1.5em
\caption{Ghost self-energy graphs.}
\label{fig:ghost}
\end{figure}
\begin{align}
-i\Sigma_\mathrm{ghost}^{ad}=&\frac{-ig^2p_E^2}{(4\pi)^2}f^{abc}f^{dbc}\int_0^{\frac{1}{2}}d\tau E_1
\left(\tau\frac{p_E^2}{\Lambda_W^2}\right)\nonumber\\
=&\frac{-ig^2p_E^2}{(4\pi)^2}f^{abc}f^{dbc}\left\{\frac{\Lambda_W^2}{p_E^2}\left[1-e^{-p_E^2/2\Lambda_W^2}\right]+\frac{1}{2}E_1\left(\frac{p_E^2}{2\Lambda_W^2}\right)\right\},
\end{align}
where
\begin{equation}
E_n(x)=\int_1^\infty e^{-xt}y^{-n}dy=\frac{1}{n-1}[e^{-x}-xE_{n-1}(x)]~~~~~~(x>0).\label{eq:expint}
\end{equation}
$E_1$ has an analytic continuation to the entire complex $p_E^2$ plane with expansion:
\begin{equation}
E_1(z)=-Ei(-z)\rightarrow E_1(z)+i\arg(z)=-\gamma-\ln|z|-\sum_{k=1}^\infty\frac{(-z)^k}{kk!}+i\arg(z),
\end{equation}
where $\gamma$ is Euler's constant, and we will take the cut along the positive $x$-axis. This self-energy does indeed have a trivial solution at $p_E^2=0$.

The non-local Lagrangian gives us the first three diagrams in Figure~\ref{fig:ghost}, but we will represent them all by the analogous local diagram since the extra vertices only produce amplitudes which are identical to their local counterparts aside from the range of Schwinger parameter integrals~\cite{Evens1991}. This is made complicated at higher orders by the presence of the measure within another process.

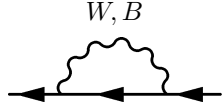
\begin{figure}
\begin{center}
\begin{fmfgraph*}(80,60)
\fmfleftn{i}{1}\fmfrightn{o}{1}
\fmf{fermion}{v1,i1}
\fmf{fermion}{o1,v2}
\fmf{fermion,tension=0.5}{v2,v1}
\fmf{boson,right,tension=0,label=$W,,B$}{v2,v1}
\end{fmfgraph*}
\end{center}
\vskip -4em
\caption{Fermion self-energy.}
\label{fig:fermion}
\end{figure}
Next we can compute the fermion self-energies (Figure~\ref{fig:fermion}) with massless fermions:
\begin{align}
-i\Sigma_\mathrm{fermion}&=\frac{2ie^2}{(4\pi)^2}\Gamma_5^+\slashed p\int_0^{\frac{1}{2}} d\tau E_1\left(\frac{\tau p_E^2}{\Lambda_W^2}\right)\nonumber\\
&=\frac{2ie^2}{(4\pi)^2}\Gamma_5^+\slashed p\left\{\frac{\Lambda_W^2}{p_E^2}\left[1-e^{-p_E^2/2\Lambda_W^2}\right]+\frac{1}{2}E_1\left(\frac{p_E^2}{2\Lambda_W^2}\right)\right\},
\end{align}
where we have used a generic coupling at the vertex:
\begin{equation}
-ie\gamma^\mu(g_v-g_a\gamma^5)
\end{equation}
and
\begin{equation}
\Gamma_5^\pm=(g_v\pm g_a\gamma^5)^2,
\label{eq:gagv}
\end{equation}
and $g_a=+1/4$ for neutrinos and up-type quarks, $g_a=-1/4$ for charged leptons and down-type quarks, while $g_v=g_a-Q\sin^2\theta_w$ where $Q$ is the fermion charge.

Again there is a pole at $\slashed p=0$. (We will not bother to specialize to each boson contribution.)

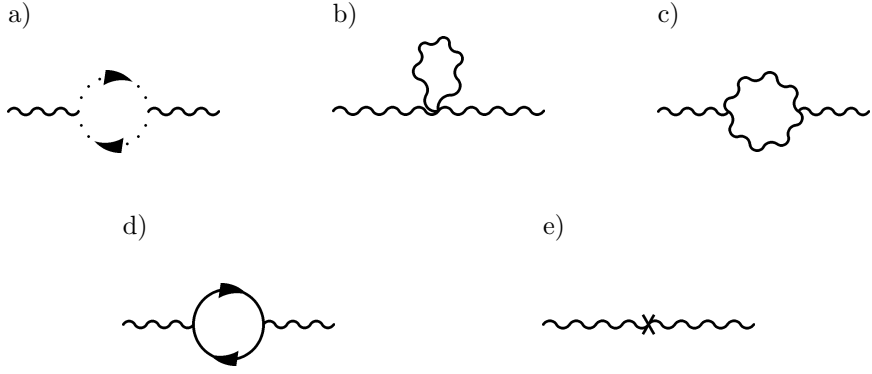
\begin{figure}
\begin{center}
\begin{minipage}{0.24\linewidth}a)\\~\\
\begin{fmfgraph*}(80,40)
\fmfleftn{i}{1}\fmfrightn{o}{1}
\fmf{boson}{v1,i1}
\fmf{boson}{o1,v2}
\fmf{dots_arrow,left,tension=0.5}{v1,v2,v1}
\end{fmfgraph*}
\end{minipage}\hskip 0.1\linewidth
\begin{minipage}{0.24\linewidth}b)\\~\\
\begin{fmfgraph*}(80,40)
\fmfleftn{i}{1}\fmfrightn{o}{1}
\fmf{boson}{i1,v1}
\fmf{boson}{v1,o1}
\fmf{boson}{v1,v1}
\end{fmfgraph*}
\end{minipage}\hskip 0.1\linewidth
\begin{minipage}{0.24\linewidth}c)\\~\\
\begin{fmfgraph*}(80,40)
\fmfleftn{i}{1}\fmfrightn{o}{1}
\fmf{boson}{i1,v1}
\fmf{boson}{v2,o1}
\fmf{boson,left,tension=0.5}{v1,v2,v1}
\end{fmfgraph*}
\end{minipage}\\\vskip 18pt
\begin{minipage}{0.24\linewidth}d)\\~\\
\begin{fmfgraph*}(80,40)
\fmfleftn{i}{1}\fmfrightn{o}{1}
\fmf{boson}{i1,v1}
\fmf{boson}{v2,o1}
\fmf{fermion,left,tension=0.5}{v1,v2,v1}
\end{fmfgraph*}
\end{minipage}\hskip 0.2\linewidth
\begin{minipage}{0.24\linewidth}e)\\~\\
\begin{fmfgraph*}(80,40)
\fmfleftn{i}{1}\fmfrightn{o}{1}
\fmf{boson}{o1,i1}
\fmf{crossed_phantom}{i1,o1}
\end{fmfgraph*}
\end{minipage}
\end{center}
\vskip -1em
\caption{Boson self-energy.}
\label{fig:boson}
\end{figure}
We can now do the gauge boson self-energies. This is where the measure first becomes important, since without it the bosons appear to have picked up an extra degree of freedom. We split the vacuum polarization tensor into longitudinal and transverse pieces,
\begin{equation}
-i\Pi_{\mu\nu}=-i\Pi^T\left(\eta_{\mu\nu}-\frac{p_\mu p_\nu}{p^2}\right)-i\Pi^L\frac{p_\mu p_\nu}{p^2}
\end{equation}
and calculate the bosonic loops in Figures~\ref{fig:boson}a-c. The ghost loop is given by
\begin{align}
-&i\Pi_\mathrm{ghost}^{Lad}=\frac{ig^2\Lambda_W^2}{(4\pi)^2}f^{abc}f^{dbc}\int_0^{\frac{1}{2}}d\tau (1-\tau)\left[\exp\left(-\frac{\tau p_E^2}{\Lambda_W^2}\right)-3\tau\frac{p_E^2}{\Lambda_W^2}E_1\left(\frac{\tau p_E^2}{\Lambda_W^2}\right)\right]\nonumber\\
=&\frac{ig^2\Lambda_W^2}{(4\pi)^2}f^{abc}f^{dbc}\left[-\frac{1}{4}\frac{p_E^2}{\Lambda_W^2}E_1\left(\frac{p_E^2}{\Lambda_W^2}\right)+\left(\frac{1}{2}-\frac{\Lambda_W^4}{p_E^4}\right)e^{-p_E^2/2\Lambda_W^2}-\frac{\Lambda^2}{2p_E^2}+\frac{\Lambda_W^4}{p_E^4}\right],
\end{align}
\begin{align}
-i\Pi_\mathrm{ghost}^{Tad}=\frac{ig^2\Lambda_W^2}{(4\pi)^2}f^{abc}f^{dbc}\int_0^{\frac{1}{2}}d\tau(1-\tau)\left[\exp\left(-\frac{\tau p_E^2}{\Lambda_W^2}\right)-\tau\frac{p_E^2}{\Lambda_W^2}E_1\left(\frac{\tau p_E^2}{\Lambda_W^2}\right)\right]&\nonumber\\
=\frac{ig^2\Lambda_W^2}{(4\pi)^2}f^{abc}f^{dbc}\left[-\frac{1}{12}\frac{p_E^2}{\Lambda_W^2}E_1\left(\frac{p_E^2}{2\Lambda_W^2}\right)+\frac{1}{6}\left(1-2\frac{\Lambda_W^2}{p_E^2}+2\frac{\Lambda_W^4}{p_E^4}\right)e^{-p_E^2/2\Lambda_W^2}\right.&\nonumber\\
\left.+\frac{\Lambda_W^2}{2p_E^2}-\frac{\Lambda_W^4}{3p_E^4}\right]&.
\end{align}
The 4-$W$ boson interaction vertex is
\begin{equation}
-i\Pi_4^{Lad}=-i\Pi_4^T=3\frac{ig^2\Lambda_W^2}{(4\pi)^2}f^{abc}f^{dbc}.
\end{equation}
The $W$-loop is given by
\begin{align}
-i\Pi_W^{Lad}=-\frac{ig^2\Lambda_W^2}{(4\pi)^2}f^{abc}f^{dbc}\int_0^{\frac{1}{2}}d\tau\left[9E_2\left(\frac{\tau p_E^2}{\Lambda_W^2}\right)-3\tau\frac{p_E^2}{\Lambda_W^2}E_1\left(\frac{\tau p_E^2}{\Lambda_W^2}\right)(4\tau(1-\tau)-1)\right]&\nonumber\\
=-\frac{ig^2\Lambda_W^2}{(4\pi)^2}f^{abc}f^{dbc}\left[-\frac{17}{16}\frac{p_E^2}{\Lambda_W^2}E_1\left(\frac{p_E^2}{\Lambda_W^2}\right)+\left(\frac{17}{8}-\frac{17}{4}\frac{\Lambda_W^2}{p_E^2}-\frac{\Lambda_W^4}{p_E^4}-18\frac{\Lambda_W^6}{p_E^6}\right)e^{-p_E^2/2\Lambda_W^2}\right.&\nonumber\\
\left.+6\frac{\Lambda_W^2}{p_E^2}-8\frac{\Lambda_W^4}{p_E^4}+18\frac{\Lambda_W^6}{p_E^6}\right],
\end{align}
\begin{align}
-i\Pi_W^{Tad}=-\frac{ig^2\Lambda_W^2}{(4\pi)^2}f^{abc}f^{dbc}\int_0^{\frac{1}{2}}d\tau\left[9E_2\left(\frac{\tau p_E^2}{\Lambda_W^2}\right)-\tau\frac{p_E^2}{\Lambda_W^2}E_1\left(\frac{\tau p_E^2}{\Lambda_W^2}\right)(2\tau(1-\tau)-5)\right]&\nonumber\\
=-\frac{ig^2\Lambda_W^2}{(4\pi)^2}f^{abc}f^{dbc}\left[-\frac{53}{96}\frac{p_E^2}{\Lambda_W^2}E_1\left(\frac{p_E^2}{\Lambda_W^2}\right)+\left(\frac{53}{48}-\frac{161}{24}\frac{\Lambda_W^2}{p_E^2}-\frac{1}{6}\frac{\Lambda_W^4}{p_E^4}-3\frac{\Lambda_W^6}{p_E^6}\right)e^{-p_E^2/2\Lambda_W^2}\right.&\nonumber\\
\left.+7\frac{\Lambda_W^2}{p_E^2}-\frac{4}{3}\frac{\Lambda_W^4}{p_E^4}+3\frac{\Lambda_W^6}{p_E^6}\right]&.
\end{align}
If we now impose gauge invariance, our $W$-boson should not have any longitudinal degrees of freedom, so
we choose as our measure:
\begin{equation}
\ln(\mu_\mathrm{inv}[W]_\mathrm{bos})=\frac{1}{2}\int d^4xW^{a\mu}\Omega_{\mu\nu}^{ab}W^{b\nu},
\end{equation}
where
\begin{align}
\Omega_{\mu\nu}^{ab}=&-\frac{ig^2\Lambda_W^2}{(4\pi)^2}f^{abc}f^{dbc}\eta_{\mu\nu}\int_0^{\frac{1}{2}}d\tau\bigg[(8+\tau)E_2\left(\frac{\tau p_E^2}{\Lambda_W^2}\right)\nonumber\\
&\hskip 2in -3\frac{p_E^2}{\Lambda_W^2}E_1\left(\frac{\tau p_E^2}{\Lambda_W^2}\right)(2\tau(1-2\tau)-1)\bigg]\nonumber\\
=&-\frac{ig^2\Lambda_W^2}{(4\pi)^2}f^{abc}f^{dbc}\eta_{\mu\nu}\bigg[\frac{5}{24}E_1\left(\frac{p_E^2}{\Lambda_W^2}\right)\nonumber\\
&\hskip 0.3in +\left(-\frac{5}{12}-\frac{31}{6}\frac{\Lambda_W^2}{p_E^2}-\frac{25}{3}\frac{\Lambda_W^4}{p_E^4}\right)e^{-p_E^2/\Lambda_W^2}+3+\frac{\Lambda_W^2}{p_E^2}+\frac{25}{3}\frac{\Lambda_W^4}{p_E^4}\bigg].
\end{align}
We are left with a purely transverse result that admits the zero mass solution at $p_E^2=0$:
\begin{align}
-i\Pi^{Tad}_\mathrm{bos}=&-\frac{ig^2\Lambda_W^2}{(4\pi)^2}f^{abc}f^{dbc}\frac{p_E^2}{\Lambda_W^2}\int_0^{\frac{1}{2}}d\tau E_1\left(\frac{\tau p_E^2}{\Lambda_W^2}\right)(8\tau(1-\tau)+2)\\
=&-\frac{ig^2\Lambda_W^2}{(4\pi)^2}f^{abc}f^{dbc}\frac{p_E^2}{\Lambda_W^2}\left[\frac{5}{3}E_1\left(\frac{p_E^2}{\Lambda_W^2}\right)\right.&\nonumber\\
&\left.-\frac{1}{3}\frac{\Lambda_W^2}{p_E^2}\left(10+4\frac{\Lambda_W^2}{p_E^2}-16\frac{\Lambda_W^4}{p_E^4}\right)e^{-p_E^2/2\Lambda_W^2}+2\frac{\Lambda_W^2}{p_E^2}+4\frac{\Lambda_W^4}{p_E^4}-\frac{16}{3}\frac{\Lambda_W^6}{p_E^6}\right].\nonumber
\end{align}

This takes care of the bosonic sector, and we now turn our attention to the fermionic sector coming from
Figure~\ref{fig:boson}d. Using the same generic coupling as above (all quantities primed at one vertex), and defining, from (\ref{eq:gagv}):
\begin{equation}
g_\pm=g_v g'_v\pm g_ag'_a,
\end{equation}
we get
\begin{align}
i\Pi_f^L=&-\frac{4iee'\Lambda_W^2}{(4\pi)^2}g_+K,\\
-i\Pi_f^T=&-\frac{4iee'\Lambda_W^2}{(4\pi)^2}g_+(K+2P),
\end{align}
where
\begin{align}
K=&~2\int_0^{\frac{1}{2}}d\tau(1-\tau)\exp\left(-\tau\frac{p_E^2}{\Lambda_W^2}\right)=-\left(\frac{\Lambda_W^2}{p_E^2}+2\frac{\Lambda_W^4}{p_E^4}\right)e^{-p_E^2/2\Lambda_W^2}+2\frac{\Lambda_W^4}{p_E^4},\\
P=&-2\frac{p_E^2}{\Lambda_W^2}\int_0^{\frac{1}{2}}d\tau\tau(1-\tau)E_1\left(\tau\frac{p_E^2}{\Lambda_W^2}\right)\nonumber\\
=&-\frac{1}{6}\frac{p_E^2}{\Lambda_W^2}E_1\left(\frac{p_E^2}{2\Lambda_W^2}\right)+\frac{1}{3}\left(1+\frac{\Lambda_W^2}{p_E^2}-4\frac{\Lambda_W^4}{p_E^4}\right)e^{-p_E^2/2\Lambda_W^2}-\frac{\Lambda_W^2}{p_E^2}+\frac{4}{3}\frac{\Lambda_W^4}{p_E^4}.
\end{align}
To rid ourselves of the longitudinal degrees of freedom, we include a measure contribution for each diagram:
\begin{equation}
\ln(\mu_\mathrm{inv}[AA']_\mathrm{ferm})=\int d^4xA^\mu\Upsilon_{\mu\nu}^{AA'}A'^\nu,
\end{equation}
\begin{equation}
\Upsilon_{\mu\nu}^{AA'}=-S\frac{4iee'\Lambda_W^2}{(4\pi)^2}g_+\eta_{\mu\nu}K,
\end{equation}
where $S=\frac{1}{2}$ if $A_\mu=A_\mu'$ and $S=1$ otherwise. This leaves just the transverse piece:
\begin{equation}
-i\Pi_f^T=-\frac{8iee'\Lambda_W^2}{(4\pi)^2}g_+P.
\end{equation}

We can specialize this to each gauge field. For the $B-B$ sector we have
\begin{align}
\Upsilon_{\mu\nu}^{BB}=&-\frac{2ig'^2\Lambda_W^2}{(4\pi)^2}\eta_{\mu\nu}K\sum_\psi((Q-T_3)^2+Q^2)=-20\frac{ig'^2\Lambda_W^2}{(4\pi)^2}\eta_{\mu\nu}K,\\
-i\Pi_{BBf}^T=&-\frac{8ig'^2\Lambda_W^2}{(4\pi)^2}P\sum_\psi((Q-T_3)^2+Q^2)=-80\frac{ig'^2\Lambda_W^2}{(4\pi)^2}K.
\end{align}
Moreover, for the $W^3-W^3$ sector, we find
\begin{align}
\Upsilon_{\mu\nu}^{W^3W^3}=&-\frac{ig^2\Lambda_W^2}{(4\pi)^2}\eta_{\mu\nu}K\sum_{q^L}1=-12\frac{ig^2\Lambda_W^2}{(4\pi)^2}\eta_{\mu\nu}K,\\
-i\Pi_{WWf}^T=&-\frac{4ig^2\Lambda_W^2}{(4\pi)^2}P\sum_{q^L}1=-48\frac{ig^2\Lambda_W^2}{(4\pi)^2}P.
\end{align}
When we diagonalize the $W^1-W^2$ sector into the physical $W^\pm$ fields, we get:
\begin{align}
\Upsilon_{\mu\nu}^{W^\pm}=&-\frac{2ig^2\Lambda_W^2}{(4\pi)^2}\eta_{\mu\nu}K\sum_{q^L}1=-24\frac{ig^2\Lambda_W^2}{(4\pi)^2}\eta_{\mu\nu}K,\\
-i\Pi_{W^\pm f}^T=&-\frac{4ig^2\Lambda_W^2}{(4\pi)^2}P\sum_{q^L}1=-48\frac{ig^2\Lambda_W^2}{(4\pi)^2}P.
\end{align}
For the $W^3-B$ mixing sector we get
\begin{align}
\label{eq:98}\Upsilon_{\mu\nu}^{WB}=&-\frac{igg'\Lambda_W^2}{(4\pi)^2}\eta_{\mu\nu}K\sum_{q^L}YT^3,\\
\label{eq:99}-i\Pi_{WBf}^T=&-\frac{2igg'\Lambda_W^2}{(4\pi)^2}P\sum_{q^L}YT^3.
\end{align}
The sum of (\ref{eq:98}) and (\ref{eq:99}) is zero in the gauge invariant case. The invariant measure is then given by the product of each piece generated above and is represented diagrammatically by Figure~\ref{fig:boson}e.

We also note that the BRST invariance implies Slavnov-Taylor identities analogous to those in the local case, which also must be satisfied for a valid perturbation theory.

\section{Symmetry Breaking}
\label{sec:symbreak}

An alternative to the standard perturbative renormalization method is to identify the vector boson self-energy with the vector boson mass $m^2_V={\Pi_\mu}^\mu(0)$. The vector boson creates a virtual fermion-anti-fermion pair which in turn creates a vector boson, producing the vector boson self-energy diagram. The fermion-anti-fermion pair can be pictured as a virtual fermion ``condensate'', which by a breaking of the fermion sector symmetry and $U(1)$ gauge invariance gives the vector boson a mass.

Let us consider the situation from a different point of view. In standard perturbation theory, we solve by successive approximations starting with the bare mass $m_{\gamma 0}$ and the bare coupling constant $e_0$ maintaining gauge invariance. However, we also entertain the idea that there are solutions which cannot be thus obtained. In fact, there exist solutions with $m_\gamma\not=0$ when the bare photon mass, $m_{\gamma0}=0$, even though the gauge symmetry forbids a finite mass $m_\gamma$. We can understand this by considering a self-consistent Hartree-Fock type of procedure~\cite{Nambu1961}. In standard perturbation theory, we compose the free and interaction parts:
\begin{equation}
L=L_0+L_I.
\end{equation}
Instead of diagonalizing $L_0$ and treating the interaction part as a perturbation, we introduce the self-energy Lagrangian $L_{\rm self}$ and split $L$ as
\begin{equation}
L=(L_0+L_{\rm self})+(L_I-L_{\rm self})=L_0'+L'_I.
\end{equation}
We can now define a new vacuum and a complete set of ``quasi-particle'' states for which each particle is an eigenmode of $L_0'$. We now solve $L_{\rm self}$ as a perturbation and determine $L_{\rm self}$ without producing additional self-energy effects. The self-consistent nature of the procedure allows the self-energy to be calculated by perturbation theory with the fields defined by a new vacuum which are already subject to the self-energy interaction.

Let us now consider a non-Abelian gauge vector field $W_\mu^a$. We assume that $W_\mu^a$ is an $SU(2)$ isospin vector which transforms as
\begin{equation}
W_\mu\rightarrow W_\mu+i\theta^a[T^a,W_\mu],
\end{equation}
with $a=1,2,3$ running over the three generators of $SU(2)$. Our action now picks up a quadratic term from the lowest order non-Abelian self-energy diagram:
\begin{equation}
g^2\Pi{\rm Tr}[T^a,T^b]W^a_\mu W^{\mu b},
\end{equation}
where $\Pi=\Pi(q^2)$ denotes the proper vector boson self-energy contribution. The gauge boson masses squared are determined by the eigenvalues of the 3 by 3 matrix $g^2\Pi{\rm Tr}[T^a,T^b]$.

Let us consider the symmetry group $G$ which is broken down to the subgroup $H$. We find that $N(G)-N(H)$ Nambu-Goldstone bosons will be generated. We start with $N(G)$ massless gauge bosons, one for each generator. Upon symmetry breaking, the $N(G)-N(H)$ Nambu-Goldstone bosons are eaten by $N(G)-N(H)$ gauge bosons, leaving $N(H)$ massless gauge bosons. For the case of $SU_L(2)\times U_Y(1)$ we have $N(G)=4$ and $N(H)=1$ and we end up with one massless gauge boson, namely, the photon. In our Lagrangian after symmetry breaking:
\begin{equation}
L_m=\frac{1}{2}g^2\Pi[T^a\cdot T^b]W^{\mu a}W_\mu^b=\frac{1}{2}W^{\mu a}(m^2)^{ab}W^b_\mu,
\end{equation}
where
\begin{equation}
(m^2)^{ab}=g^2\Pi[T^a\cdot T^b]
\end{equation}
denotes the mass matrix.

We now diagonalize $(m^2)^{ab}$ to obtain the masses of the gauge bosons. A calculation of the eigenvectors determines the combination of eigenstates for the masses. The mass matrix $(m^2)^{ab}$ is a 4 by 4 matrix with 1 zero eigenvalue for our group $SU_L(2)\times U_Y(1)$. Since $U(1)$ remains unbroken by our symmetry breaking mechanism, the generator $T^c$ associated with the $U(1)$ symmetry satisfies $T^c\Pi=0$, leaving the photon massless. We have
\begin{eqnarray}
L_m&=&\frac{1}{4}g^2\Pi W^+_\mu W^{-\mu}+\frac{1}{8}\Pi(gW^3_\mu-g'B_\mu)^2.\label{massLag}
\end{eqnarray}
We get
\begin{equation}
Z_\mu=c_wW_\mu^3-s_wB_\mu
\end{equation}
describing the neutral $Z$ boson, while the photon is described by
\begin{equation}
A_\mu=s_wW_\mu^3+c_wB_\mu.
\end{equation}
We have $m_Z^2=\Pi(g^2+g^{'2})/4$ giving
\begin{equation}
\rho=\frac{m_W^2}{m_Z^2c_w^2}=1,
\end{equation}
which is the standard tree graph result.

Let us introduce the spin-1 vector $V^\alpha_\mu$ and from the loop graph in Figure \ref{fig:boson}d, we obtain the mass matrix:
\begin{equation}
{\cal M}=V^\alpha_\mu m^2_{\alpha\beta}V ^{\mu\beta},\quad (\alpha,\beta=a,0),
\end{equation}
where $V^a_\mu=W_\mu^a$ and $V^0_\mu=B_\mu$. The most general form of the spin-1 vector boson mass matrix that correctly gives the symmetry-breaking pattern $SU_L(2)\times U_Y(1)\rightarrow U_{\rm em}(1)$ is given by~\cite{Burgess2007}:
\begin{equation}
m^2_{\alpha\beta}=\begin{pmatrix}
m_W^2&&&\\
&m_W^2&&\\
&&m_3^2&m^2\\
&&m^2&m_0^2
\end{pmatrix}.
\end{equation}
The unbroken electromagnetic gauge invariance that guarantees a massless photon dictates that the upper left $2\times 2$ block of the matrix be proportional to the unit matrix: $m_W^2I_{2\times 2}$. Moreover, it also says that the upper-right and the lower-left blocks must vanish. The vanishing of one of the eigenvalues guarantees a massless photon, which corresponds to:
\begin{equation}
{\rm det}\begin{pmatrix}
m_3^2&m^2\\
m^2&m_0^2.
\end{pmatrix}=m_3^2m_0^2-m^4=0.
\end{equation}
Eliminating $m_0^2$ in favor of $\theta_w$ by using the relation
\begin{equation}
\tan\theta_w=\frac{m^2}{m^2_3}=\left|\frac{m_0}{m_3}\right|,
\end{equation}
we obtain the non-zero eigenvalue:
\begin{equation}
m_Z^2=\mathrm{tr}
\begin{pmatrix}
m_3^2&m^2\\
m^2&m_0^2
\end{pmatrix}=m_0^2+m_3^2=m_3^2(1+\tan^2\theta_w)=m_3^2\sec^2\theta_w.
\end{equation}
We now arrive at the relation:
\begin{equation}
\frac{m_W}{m_Z}=\frac{g}{\sqrt{g^2+g^{'2}}}=c_w.
\end{equation}

\section{Breaking the Symmetry with a Path Integral Measure}
\label{sec:measure}

It has been recognized that quantization can break classical symmetries. In particular, classical symmetries can be broken through the choice of measure and the associated Jacobian transformations \cite{Fujikawa2004}. Two important historical cases of symmetry breaking by quantization of the measure are the chiral anomaly and the Weyl or conformal anomaly.

In our case, we follow a similar route. We break $SU_L(2)\times U_Y(1)$ down to $U_\mathrm{em}(1)$ not at the classical level as is done in the standard model, which generates boson masses at tree level, but in the quantum regime \cite{Moffat1991,Clayton1991b,Moffat2007f}, so that all the effects show up at loop order (which is where the non-locality shows up as well, as both are quantum effects). This means leaving the action invariant and modifying the measure, which alters the quantization of the theory, in order to produce the desired results.

Even though the choice of the symmetry breaking measure is not unique, after an initial {\it ansatz} chosen as the minimal scheme, the rest of the method follows directly. This is no worse than the standard model with a Higgs mechanism, where it is assumed that the minimal spontaneous symmetry breaking is assumed to be caused by an isospin doublet scalar field. An alternative would be to assume that the scalar field transforms as an isotriplet or as an isodoublet and an isotriplet, but this would yield the incorrect answer for the $W$ and $Z$ masses. Thus the minimal choice for the symmetry breaking measure in the path integral is no more {\it ad hoc} than the choice of symmetry breaking in the standard Higgs motivated model. We allow the fermion mass generation mechanism to come in to effect, and so we work with massive fermions.

The symmetry breaking measure in our path integral generates three new degrees of freedom as scalar Nambu-Goldstone bosons that give the $W^\pm$ and $Z^0$ bosons longitudinal modes, which makes them massive while retaining a massless photon.

Since we want to mix the $W^3$ and $B$ to get a massive $Z$ and a photon, we need to work with the measure in a sector which is common to all gauge bosons. This implies working with the fermion contributions and leaving the bosonic and ghost contributions invariant. We shall take it as given that the fermions have acquired a mass, generated by the mechanism described in the following section. The self-energy contribution coming from Figure~\ref{fig:boson}d looks like:
\begin{eqnarray}
-i\Pi_f^L&=&-\frac{4iee'\Lambda_W^2}{(4\pi)^2}[g_+(K_{m_1m_2}-L_{m_1m_2})+g_-M_{m_1m_2}],\\
-\Pi_f^T&=&-\frac{4iee'\Lambda_W^2}{(4\pi)^2}[g_+(K_{m_1m_2}-L_{m_1m_2}+2P_{m_1m_2})+g_-M_{m_1m_2}],
\end{eqnarray}
where we define
\begin{align}
K_{m_1m_2}=&\int_0^{\frac{1}{2}}d\tau(1-\tau)\left[\exp\left(-\tau\frac{p_E^2}{\Lambda_W^2}-f_{m_1m_2}\right)+\exp\left(-\tau\frac{p_E^2}{\Lambda_W^2}-f_{m_2m_1}\right)\right],
\end{align}
\begin{align}
\label{eq:P}P_{m_1m_2}=&-\frac{p_E^2}{\Lambda_W^2}\int_0^{\frac{1}{2}}d\tau\tau(1-\tau)\left[E_1\left(\tau\frac{p_E^2}{\Lambda_W^2}+f_{m_1m_2}\right)+E_1\left(\tau\frac{p_E^2}{\Lambda_W^2}+f_{m_2m_1}\right)\right],
\end{align}
\begin{align}
L_{m_1m_2}=&\int_0^{\frac{1}{2}}d\tau(1-\tau)\left[f_{m_1m_2}E_1\left(\tau\frac{p_E^2}{\Lambda_W^2}+f_{m_1m_2}\right)
+f_{m_2m_1}E_1\left(\tau\frac{p_E^2}{\Lambda_W^2}+f_{m_2m_1}\right)\right],
\end{align}
\begin{align}
M_{m_1m_2}=&\frac{m_1m_2}{\Lambda_W^2}\int_0^{\frac{1}{2}}d\tau\left[E_1\left(\tau\frac{p_E^2}{\Lambda_W^2}
+f_{m_1m_2}\right)+E_1\left(\tau\frac{p_E^2}{\Lambda_W^2}+f_{m_2m_1}\right)\right],
\end{align}
and where
\begin{equation}
f_{m_1m_2}=\frac{m_1^2}{\Lambda_W^2}+\frac{\tau}{1-\tau}\frac{m_2^2}{\Lambda_W^2}.
\end{equation}
If we insert this into the quadratic terms in the action and invert, we get the corrected propagator (in a general gauge):
\begin{equation}
iD^{\mu\nu}=-i\left(\frac{\eta^{\mu\nu}-\frac{p^\mu p^\nu}{p^2}}{p^2-\Pi_f^T}+\frac{\frac{\xi p^\mu p^\nu}{p^2}}{p^2-\xi\Pi_f^L}\right),
\label{eq:vecprop}
\end{equation}
and when the longitudinal piece is nonzero in the unitary gauge (where only the physical particle spectrum remains), we have no unphysical poles in the longitudinal sector. In this way, we can assure ourselves that we are not introducing spurious degrees of freedom into the theory.

In the diagonalized $W^\pm$ sector, we get
\begin{align}
-i\Pi_{W^\pm f}^L=&-\frac{ig^2\Lambda_W^2}{(4\pi)^2}\sum_{q^L}(K_{m_1m_2}-L_{m_1m_2}),\\
-i\Pi_{W^\pm f}^T=&-\frac{ig^2\Lambda_W^2}{(4\pi)^2}\sum_{q^L}(K_{m_1m_2}-L_{m_1m_2}+2P_{m_1m_2}).
\label{eq:WWPi}
\end{align}

We note that at $p^2=0$,
\begin{equation}
-i\Pi_{W^\pm f}^L\bigg|_{p^2=0}=-i\Pi_{W^\pm f}^T\bigg|_{p^2=0}=-\frac{ig^2\Lambda_W^2}{(4\pi)^2}\sum_{q^L}(K_{m_1m_2}-L_{m_1m_2})
\bigg|_{p^2=0}\ne 0.
\end{equation}
This introduces three Nambu-Goldstone degrees of freedom into the $W-B$ sector and the vector bosons acquire a longitudinal part and a corresponding mass.

We go on to calculate the self-energy in the $W^3$ sector as
\begin{align}
-i\Pi_{W^3f}^L=&-\frac{1}{2}\frac{ig^2\Lambda_W^2}{(4\pi)^2}\sum_\psi(K_{mm}-L_{mm}),\\
-i\Pi_{W^3f}^T=&-\frac{1}{2}\frac{ig^2\Lambda_W^2}{(4\pi)^2}\sum_\psi(K_{mm}-L_{mm}+2P_{mm}).
\end{align}
It is clear that if we want the $B$ sector to mix with this, we need to make the vacuum polarization tensor look very similar. This is what motivates the choice of symmetry breaking measure, after one makes the initial
{\it ansatz}. In the $B$ sector we have
\begin{align}
-i\Pi_{Bf}^L=&-\frac{1}{2}\frac{ig'^2\Lambda_W^2}{(4\pi)^2}\sum_\psi[16(Q-T^3)^2(K_{mm}-L_{mm})+32Q(Q-T^3)M_{mm}],\\
-i\Pi_{Bf}^T=&-\frac{1}{2}\frac{ig'^2\Lambda_W^2}{(4\pi)^2}\sum_\psi[16(Q-T^3)^2(K_{mm}-L_{mm}+2P_{mm})+32Q(Q-T^3)M_{mm}],
\end{align}
so we write the measure contribution as
\begin{equation}
\Upsilon_{\mu\nu}^{BB}=-\frac{ig'^2\Lambda_W^2}{(4\pi)^2}\eta_{\mu\nu}\sum_\psi\left[\left(\frac{1}{2}-8(Q-T^3)^2
\right)(K_{mm}-L_{mm})-16Q(Q-T^3)M_{mm}\right]
\end{equation}
and we are then left with
\begin{eqnarray}
-i\Pi_{Bf}^L&=&-\frac{1}{2}\frac{ig'^2\Lambda_W^2}{(4\pi)^2}\sum_\psi(K_{mm}-L_{mm}),\\
-i\Pi_{Bf}^T&=&-\frac{1}{2}\frac{ig'^2\Lambda_W^2}{(4\pi)^2}\sum_\psi[(K_{mm}-L_{mm})+32(Q-T^3)^2P_{mm}].
\end{eqnarray}
Note that the pieces that contribute to the mass generation are identical to those given above. The presence of the extra piece proportional to $p^2$ will not give any problems in the mass matrix, and will produce a $Z$-photon mixing that contains no extra poles. The $B-W^3$ mixing sector originally looks like
\begin{align}
-i\Pi_{W^3Bf}^L=&-\frac{4igg'\Lambda_W^2}{(4\pi)^2}\sum_\psi[T^3(Q-T^3)(K_{mm}-L_{mm})+QM_{mm}],\\
-i\Pi_{W^3Bf}^T=&-\frac{4igg'\Lambda_W^2}{(4\pi)^2}\sum_\psi[T^3(Q-T^3)(K_{mm}-L_{mm}+2P_{mm})+QM_{mm}].
\end{align}
Thus, to make the mass contributions look identical, we write
\begin{equation}
\Upsilon_{\mu\nu}^{W^3B}=-\frac{igg'\Lambda_W^2}{(4\pi)^2}\eta_{\mu\nu}\sum_\psi\left[\left(-\frac{1}{2}-4T^3(Q-T^3)\right)(K_{mm}-L_{mm})-4QM_{mm}\right].
\end{equation}
Then we have
\begin{eqnarray}
-i\Pi_{W^3B f}^L&=&\frac{1}{2}\frac{igg'\Lambda_W^2}{(4\pi)^2}\sum_\psi(K_{mm}-L_{mm}),\\
-i\Pi_{W^3B f}^T&=&\frac{1}{2}\frac{igg'\Lambda_W^2}{(4\pi)^2}\sum_\psi[(K_{mm}-L_{mm})-8T^3(Q-T^3)P_{mm}].
\end{eqnarray}

We can now write the new fields, defined by the transformation in (\ref{eq:2.35}), and find that only the diagonal $Z-Z$ piece has a longitudinal part
\begin{equation}
-i\Pi_{Zf}^L=-\frac{1}{2}\frac{i(g^2+g'^2)\Lambda_W^2}{(4\pi)^2}\sum_\psi(K_{mm}-L_{mm}).
\end{equation}
In the transverse sector things look a bit more complicated. For the $Z-Z$ part we get
\begin{align}
-i\Pi_{Zf}^T=&-\frac{1}{2}\frac{i(g^2+g'^2)\Lambda_W^2}{(4\pi)^2}\nonumber\\
&\times\sum_\psi[(K_{mm}-L_{mm})+P_{mm}(2c_w^4+s_w^432(Q-T^3)^2-16s_w^2c_w^2T^3(Q-T^3))].
\label{eq:ZZPi}
\end{align}
The pure photon sector gives
\begin{align}
-i\Pi_{Af}^T=&-\frac{1}{2}\frac{i(g^2+g'^2)\Lambda_W^2}{(4\pi)^2}c_w^2s_w^2\nonumber\\
&\times\sum_\psi P_{mm}(2+32(Q-T^3)^2+16T^3(Q-T^3)).
\label{eq:photon}
\end{align}
We observe from (\ref{eq:photon}) that $\Pi_A^T(0)=0$, as follows from (\ref{eq:P}), guaranteeing a massless photon.

Finally we obtain for the mixing sector:
\begin{align}
-i\Pi_{AZf}^T=&-\frac{1}{2}\frac{i(g^2+g'^2)\Lambda_W^2}{(4\pi)^2}c_w^2s_w^2\nonumber\\
&\times\sum_\psi P_{mm}[2c_w^2-32s_w^2(Q-T^3)^2-16T^3(Q-T^3)(s_w^2-c_w^2)].
\end{align}

To calculate boson masses, we note the form of the massive vector boson propagator (\ref{eq:vecprop}). When we consider the scattering of longitudinally polarized vector bosons, the terms containing $p_\mu p_\nu$ cancel out. In the remaining term, $\Pi_f^T$ appears in the same place where, in the standard model, $m_V^2$ is present. We therefore make the identification
\begin{equation}
m_V^2=\Pi_f^T.
\end{equation}
This allows us to calculate the masses of the $W^\pm$ and $Z^0$ bosons or conversely, use their experimentally known masses to calculate $\Lambda_W$, which we demonstrate later in section~\ref{sec:rho}.

The boson masses we obtained are running \cite{Moffat2008c}, and suppressed at high energy. We find that, at high energies, $\Pi_f(p^2)\propto p^{-4}$. While this suppression is sufficient to ensure that the theory does not violate unitarity \cite{Moffat2008c}, it is polynomial in nature. Therefore, we conclude that {\it the mass degrees of freedom never vanish at high energy}.

\section{Fermion masses}
\label{sec:fermass}

In earlier work, we derived fermion masses from adding an $SU_L(2)\times U_Y(1)$ invariant four-fermion interaction to our electroweak model Lagrangian~\cite{Moffat1991,Clayton1991b}. However, following the derivation of fermion masses in ref.~\cite{Moffat2007f} we will generate fermion masses from the finite one-loop fermion self-energy graph. The one-loop self-energy graphs are shown in Figure \ref{fig:fermion}. This method of deriving fermion masses is more economical in assumptions, as we obtain the masses from our original massless electroweak Lagrangian by calculating fermion self-energy graphs.

A fermion particle obeys the equation:
\begin{equation}
\label{Dirac}
\slashed p-m_{0f}-\Sigma(p)=0,
\end{equation}
for
\begin{equation}
\label{Dirac2}
\slashed p-m_f=0.
\end{equation}
Here, $m_{0f}$ is the bare fermion mass, $m_f$ is the observed fermion mass and $\Sigma(p)$ is the finite proper self-energy part. We have
\begin{equation}
m_f-m_{0f}=\Sigma(p,m_f,g,\Lambda_f)\vert_{\slashed p-m_f=0},
\end{equation}
where $\Lambda_f$ denotes the energy scales for lepton and quark masses.

A solution of (\ref{Dirac}) and (\ref{Dirac2}) can be found by successive approximations starting from the bare mass $m_{0f}$.

The one-loop correction to the self-energy of a fermion with mass $m_f$ in the regularized theory of the electromagnetic field can be written as \cite{Evens1991}:
\begin{equation}
-i\Sigma(p)=i\int\frac{d^4k}{(2\pi)^4}ie\gamma_\mu\frac{i}{\slashed{p}-\slashed{k}-m_f}ie\gamma_\nu\frac{-i\eta^{\mu\nu}}{k^2}\exp\left(\frac{p^2-m_f^2}{\Lambda_f^2}+\frac{(p-k)^2-m_f^2}{\Lambda_f^2}+\frac{k^2}{\Lambda_f^2}\right).
\end{equation}
When a massive vector boson is present with mass $m_V$, the self-energy correction reads
\begin{align}
-&i\Sigma(p)=\\
&i\int\frac{d^4k}{(2\pi)^4}ig\gamma_\mu\frac{i}{\slashed{p}-\slashed{k}-m_f}ig\gamma_\nu\frac{-i\eta^{\mu\nu}}{k^2-m_V^2}\exp\left(\frac{p^2-m_f^2}{\Lambda_f^2}+\frac{(p-k)^2-m_f^2}{\Lambda_f^2}+\frac{k^2-m_V^2}{\Lambda_f^2}\right)\nonumber\\
=&\frac{-ig^2}{8\pi^4}\exp\left(\frac{p^2-m_f^2}{\Lambda_f^2}\right)\int d^4k\frac{-\slashed{p}+\slashed{k}+2m_f}{(p-k)^2-m_f^2}\frac{1}{k^2-m_V^2}\exp\left(\frac{(p-k)^2-m_f^2}{\Lambda_f^2}+\frac{k^2-m_V^2}{\Lambda_f^2}\right),\nonumber
\end{align}
where $g$ is the appropriate coupling constant, and we made use of the identity $\gamma_\mu\slashed{p}=\gamma_\mu\gamma_\nu p^\nu=(2\eta_{\mu\nu}-\gamma_\nu\gamma_\mu)p^\nu=2p_\mu-\slashed{p}\gamma_\mu$. Promoting the propagator to Schwinger (proper time) integrals using
\begin{equation}
-\frac{1}{k^2-m^2}=\int_1^\infty\frac{d\tau}{\Lambda^2}\exp\left((\tau-1)\frac{k^2-m^2}{\Lambda^2}\right),
\end{equation}
we obtain
\begin{align}
\Sigma(p)&=\frac{g^2}{8\pi^4}\exp\left(\frac{p^2-m_f^2}{\Lambda_f^2}\right)\int_1^\infty\frac{d\tau_1}{\Lambda_f^2}\int_1^\infty\frac{d\tau_2}{\Lambda_f^2}\int d^4k(-\slashed{p}+\slashed{k}+2m_f)\nonumber\\
&\times\exp\left(\tau_1\frac{(p-k)^2-m_f^2}{\Lambda_f^2}+\tau_2\frac{k^2-m_V^2}{\Lambda_f^2}\right).
\end{align}
The exponential term can be completed to a full square by shifting the integration variable to $q=k-\tau_1/(\tau_1+\tau_2)p$. Thereafter, we get
\begin{align}
\Sigma(p)&=\frac{g^2}{8\pi^2}\exp\left(\frac{p^2-m_f^2}{\Lambda_f^2}\right)\int_1^\infty d\tau_1\int_1^\infty d\tau_2\left(\frac{-\tau_2}{(\tau_1+\tau_2)^3}\slashed{p}+\frac{2}{(\tau_1+\tau_2)^2}m_f\right)\nonumber\\
&\times\exp\left(\frac{\tau_1\tau_2}{\tau_1+\tau_2}\frac{p^2}{\Lambda_f^2}-\tau_1\frac{m_f^2}{\Lambda_f^2}-\tau_2\frac{m_V^2}{\Lambda_f^2}\right).
\end{align}
We note that $\Sigma(p)$ has exponential behavior at large $p$. At $p=0$, this integral becomes
\begin{equation}
\Sigma(0)=\frac{g^2}{4\pi^2}\exp\left(\frac{-m_V^2}{\Lambda_f^2}\right)m_f\left[E_1\left(\frac{2m_f^2}{\Lambda_f^2}\right)-\frac{m_V^2}{\Lambda_f^2}\int\limits_2^\infty d\tau\exp\left(\tau\frac{m_V^2-m_f^2}{\Lambda_f^2}\right)E_1\left(\tau\frac{m_V^2}{\Lambda_f^2}\right)\right],
\end{equation}
where the exponential integral $E_1$ was defined in (\ref{eq:expint}).

We now identify the fermion mass as $m_f=\Sigma(0)$:
\begin{equation}
m_f=\frac{g^2}{4\pi^2}\exp\left(\frac{-m_V^2}{\Lambda_f^2}\right)m_f\left[E_1\left(\frac{2m_f^2}{\Lambda_f^2}\right)-\frac{m_V^2}{\Lambda_f^2}\int\limits_2^\infty d\tau\exp\left(\tau\frac{m_V^2-m_f^2}{\Lambda_f^2}\right)E_1\left(\tau\frac{m_V^2}{\Lambda_f^2}\right)\right].
\label{eq:mf}
\end{equation}
In addition to admitting a trivial solution at $m_f=0$, this equation also has non-trivial solutions that can be computed numerically. In a theory with a single massless vector boson, (\ref{eq:mf}) can be expressed in closed form, and we get
\begin{equation}
m_f=\frac{g^2}{4\pi^2}m_fE_1\left(\frac{2m_f^2}{\Lambda_f^2}\right).
\label{eq:mfsol}
\end{equation}
This equation is also valid approximately when $m_V\ll\Lambda_f$, as the second term inside the square brackets in (\ref{eq:mf}) becomes small. A solution to (\ref{eq:mfsol}) is obtained when
\begin{equation}
\frac{m_f}{\Lambda_f}=\sqrt{\frac{1}{2}E^{-1}_1\left(\frac{4\pi^2}{g^2}\right)}.
\end{equation}
Using the electroweak coupling constant $g\simeq 0.649$, we obtain
\begin{equation}
\Lambda_f\simeq 4.3\times10^{20}m_f.
\end{equation}
For quarks, we use the strong coupling constant $g_s\simeq 1.5$, and also introduce a color factor 3. Thereafter, we obtain
\begin{equation}
\Lambda_f\simeq 35m_f.
\end{equation}
For a top quark mass $m_t=171.2$~GeV, the corresponding energy scale is about $\Lambda_t\simeq 6$~TeV.

In these calculations, $\Lambda_f$ plays a role that is similar to that of the diagonalized fermion mass matrix in the standard model. The number of undetermined parameters, therefore, is the same as in the standard model: for each fermion, a corresponding $\Lambda_f$ determines its mass.

Our model permits massive neutrinos. However, as the $\Lambda_f$ correspond to the diagonal components of a fermion mass matrix, off-diagonal terms are absent, and no flavor mixing takes place. Therefore, self-energy calculations alone are not sufficient to account for observed neutrino oscillations.

However, in addition to fermion self-energy graphs, another case must be considered. Emission or absorption of a charged vector boson $W^\pm$ can be flavor violating, through the off-diagonal components of the CKM matrix. In the standard model, such flavor violating terms are not considered significant, due to the smallness of the corresponding CKM matrix elements. However, in our regularized theory, additional factors $\Lambda_{ff'}$ enter into the picture in a manner similar to the self-energy calculation we just described. These may include terms that correspond to the off-diagonal elements of the neutrino mass matrix, offering a natural explanation for neutrino oscillations without having to introduce new interactions.

\section{Calculation of the $\rho$ parameter and $\Lambda_W$}
\label{sec:rho}

When we consider the scattering of longitudinally polarized vector bosons, the vector boson propagator (\ref{eq:vecprop}) reads
\begin{equation}
iD^{\mu\nu}(p^2)=\frac{-i\eta^{\mu\nu}}{p^2-\Pi_f^T(p^2)},
\end{equation}
where we explicitly indicated the dependence of the self-energy and the propagator on momentum. This differs from the vector boson propagator of the standard model in that the squared mass $m_V^2$ of the vector boson is replaced by the self-energy term $\Pi_f^T$. For an on-shell vector boson, demanding agreement with the standard model requires that the following consistency equation be satisfied:
\begin{equation}
m_V^2=\Pi_f^T(m_V^2).
\label{eq:vecmass}
\end{equation}
For the $Z$-boson, the on-shell mass $m_Z$ is well known from experiment. The right-hand side of (\ref{eq:vecmass}) is determined by (\ref{eq:ZZPi}), and we find that it contains terms that include the electroweak coupling constant, the Weinberg angle, fermion masses, and the $\Lambda_W$ parameter. As all these except $\Lambda_W$ are known from experiment, the equation
\begin{equation}
m_Z^2=\Pi^T_{Zf}(m_Z^2),
\end{equation}
the right-hand side of which contains $\Lambda_W$ through (\ref{eq:ZZPi}), can be used to determine $\Lambda_W$. Using the values
\begin{eqnarray}
g&=&0.649,\\
\sin^2\theta_w&=&0.2312,\\
m_t&=&171.2~\mathrm{GeV},
\end{eqnarray}
(the calculation is not sensitive to the much smaller masses of the other 11 fermions), we get
\begin{equation}
\Lambda_W=541.9~\mathrm{GeV},
\end{equation}
where the precision of $\Lambda_W$ is determined by the precision to which the $Z$-mass is known, $m_Z=91.1876\pm 0.0021$~GeV \cite{PDG2008}, and it is not sensitive to the lack of precision knowledge of the top quark mass or the other quark masses. Knowing $\Lambda_W$ allows us to solve the consistency equation for the $W$-boson mass. Treating $m_W$ as unknown, we solve using (\ref{eq:WWPi}),
\begin{equation}
m_W^2=\Pi^T_{Wf}(m_W^2),
\end{equation}
for $m_W$, and obtain
\begin{equation}
m_W\simeq 80.05~\mathrm{GeV}.
\end{equation}
This result, which does not incorporate radiative corrections, is actually slightly closer to the experimental value $m_W=80.398\pm 0.025$~GeV \cite{PDG2008} than the comparable tree-level standard model prediction $m_W\simeq 79.95$~GeV, obtained using $\rho=1$. This is anticipated as our regularization scheme will introduce some suppression of higher-order corrections at the energy scale of $m_W$. We have not yet carried out these calculations.

This mass estimate also leads to a non-trivial prediction of the $\rho$ parameter. Using the definition
\begin{equation}
\rho=\frac{m_W^2}{m_Z^2\cos^2\theta_w},
\end{equation}
we get
\begin{equation}
\rho\simeq 1.0023.
\end{equation}

\section{Conclusions and Outlook}

An electroweak model without a Higgs particle that breaks $SU_L(2)\times U_Y(1)$ has been developed, based on a finite quantum field theory. We begin with a massless and gauge invariant theory that is UV complete, Poincar\'e invariant and unitary to all orders of perturbation theory. A fundamental energy scale $\Lambda_W$ enters into the calculations of the finite Feynman loop diagrams. A path integral is formulated that generates all the Feynman diagrams in the theory. The self-energy boson loop graphs with internal fermions comprised of the observed 12 quarks and leptons have an associated measure in the path integral that is broken to generate 3 Nambu-Goldstone scalar modes that give the $W^\pm$ and the $Z^0$ bosons masses, while retaining a zero mass photon.

It is shown in a separate article~\cite{Moffat2008c} that the $W_LW_L\rightarrow W_LW_L$ and $e^+e^-\rightarrow W^+_LW^-_L$ amplitudes do not violate unitarity at the tree graph level due to the running with energy of the electroweak coupling constants $g, g'$ and $e$. This is essential for the physical consistency of the model as is the case in the standard Higgs electroweak model. A self-consistent calculation of the energy scale yields $\Lambda_W=542$ GeV and a prediction of the W mass from the $W$-boson self-energy diagrams in the symmetry broken phase gives $m_W=80.05$ GeV, which is accurate to $0.5\%$. This calculation has to be improved by including radiative corrections, but the accuracy of this first-order prediction for the $W$ mass is encouraging. A calculation of the $\rho$ parameter yields $\rho=1.0023$ and this calculation must also be repeated to include radiative corrections; this result can be compared to the standard Higgs EW model at tree level, $\rho=1$.

The unitary tree level amplitudes differ at higher energies compared to the standard model and this will allow the Higgsless and standard EW models to be distinguished from one another at the LHC.

There is no hierarchy problem in the Higgsless FEW, so the model does not require any new particles to be detected at the LHC to resolve this long-standing problem. We find that it is possible to include neutrino masses as is required by experiment in an economical way via the fermion mass generation mechanism. The fermion masses in the Higgsless model are generated by the fermion self-energy diagrams through a self-consistent mass gap equation, which also determines the neutrino masses with fundamental energy scales $\Lambda_\nu$. For the top quark mass, $m_t=171.2$ GeV the corresponding energy scale is, $\Lambda_t \sim 6$ TeV. We can produce neutrino flavor mixing through a mass matrix with off-diagonal energy scales $\Lambda_{ff'}$. This fits naturally into the quantum loop mass generation mechanism as a new way to interpret neutrino oscillation experiments.

\end{fmffile}

\section*{Acknowledgements}

We thank Michael Clayton for providing us with helpful research input during the preparation of this paper. Portions of this paper have previously appeared in his thesis \cite{Clayton1991a}. This work was supported by the Natural Sciences and Engineering Research Council of Canada. Research at the Perimeter Institute for Theoretical Physics is supported by the Government of Canada through NSERC and by the Province of Ontario through the Ministry of Research and Innovation (MRI).

\bibliography{refs}

\end{document}